\documentclass[journal,onecolumn]{IEEEtran}
\ifCLASSINFOpdf
\else
\fi
\newcommand{\F}{\mathbb{F}}

\newcommand {\ccc}{{\mathbf{c}}}
\newcommand {\xxx}{{\mathbf{x}}}
\newcommand {\yyy}{{\mathbf{y}}}
\newcommand{\C}{{\mathcal{C}}}

\newcommand{\SSS}{{\mathcal{S}}}
\newcommand{\GRS}{{\mathrm{GRS}}}

\newcommand{\BBB}{{\mathcal{B}}}

\newcommand{\HZ}{{\mathrm{HZ}}}
\newcommand{\uuu}{{{\mathbf{u}}}}

\newcommand{\vvv}{{{\mathbf{v}}}}

\newcommand{\supp}{{\mathrm{supp}}}
\newcommand{\wt}{{{\rm{wt}}}}
\newcommand{\rank}{{\rm rank}}

\makeatletter

\newcommand{\Rmnum}[1]{\expandafter\@slowromancap\romannumeral #1@}
\makeatother

\usepackage[colorlinks, linkcolor=red, citecolor=blue]{hyperref}
\usepackage{amsmath}
\usepackage{amssymb}
\usepackage{bm}
\usepackage{amssymb}
\usepackage{amsthm}
\usepackage{multirow,booktabs}
\usepackage{cases}
\usepackage{tabularx}
\usepackage{adjustbox}
\usepackage[figuresright]{rotating}
\usepackage{longtable}
\usepackage{booktabs}
\usepackage{multirow}
\usepackage{color}
\usepackage{cite}
\usepackage{multirow,booktabs}
\usepackage{cases}
\usepackage{tabularx}
\usepackage{adjustbox}
\usepackage[figuresright]{rotating}
\usepackage{longtable}
\usepackage{booktabs}
\usepackage{multirow}
\usepackage{color}
\usepackage{lineno,hyperref,mathtools}
\usepackage{soul}
\usepackage{threeparttable}
\usepackage{booktabs}
\usepackage{pifont}
\usepackage{pgfplots}
\usepackage{amsmath}
\usepackage{threeparttable}
\pgfplotsset{compat=1.18}
\allowdisplaybreaks   
\theoremstyle{definition} 
\newtheorem{theorem}{Theorem}

\newtheorem{problem}{Open Problem}
\newtheorem{lemma}[theorem]{Lemma}

\theoremstyle{definition} 
\newtheorem{remark}{Remark}

\newtheorem{definition}{Definition}

\begin{document}
\begin{sloppypar}
%

\title{On optimal quantum LRCs from the Hermitian construction and $t$-designs}
\author{Yang Li, Shitao Li, Huimin Lao, Gaojun Luo, San Ling 
\thanks{
    This research was supported by Nanyang Technological University under Research Grant 04INS000047C230GRT01 and
    the Anhui Provincial Natural Science Foundation under Grant 2408085MA014.} 
\thanks{Yang Li, Huimin Lao, and San Ling are with the School of Physical and Mathematical Sciences, 
Nanyang Technological University, 21 Nanyang Link, Singapore 637371, Singapore.
(Emails: li-y@ntu.edu.sg, huimin.lao@ntu.edu.sg, and lingsan@ntu.edu.sg)}
\thanks{San Ling is also with VinUniversity, Vinhomes Ocean Park, Gia Lam, Hanoi, Vietnam. (Email: ling.s@vinuni.edu.vn)}
\thanks{Shitao Li is with the School of Internet, Anhui University, Hefei, Anhui
230039, China. (Email: lishitao0216@163.com)}
\thanks{Gaojun Luo is with the School of Mathematics, Nanjing University of Aeronautics and Astronautics, 
Nanjing, Jiangsu 211106, China. (Email: gaojun\_luo@nuaa.edu.cn)}
}

%
%
%

%
%

\markboth{}%
{Shell \MakeLowercase{\textit{et al.}}: Bare Demo of IEEEtran.cls for Journals}
%



\maketitle

\begin{abstract}
In a recent work, quantum locally recoverable codes (qLRCs) have been introduced for their potential application in large-scale quantum data storage and implication for quantum LDPC codes. 
This work focuses on the bounds and constructions of qLRCs derived from the Hermitian construction, 
which solves an open problem proposed by Luo $et~al.$ (IEEE Trans. Inf. Theory, 71(3): 1794–1802, 2025).
We present four bounds for qLRCs and give comparisons in terms of their asymptotic formulas. 
We construct several new infinite families of NMDS codes, with general and flexible dimensions, 
that support $t$-designs for $t \in \{2, 3\}$, and apply them to obtain Hermitian dual-containing classical LRCs (cLRCs). 
As a result, we derive three explicit families of optimal qLRCs.  
Compared to the known qLRCs obtained by the CSS construction, our optimal qLRCs offer new and more flexible parameters. 
It is also worth noting that the constructed cLRCs themselves are interesting as they are optimal with respect to four distinct bounds for cLRCs. 
\end{abstract}


%
\IEEEpeerreviewmaketitle

\section{Introduction}\label{sec.introduction}

Let $\F_q$ denote the {\em finite field} of size $q$, and let $\F_q^*=\F_q\setminus \{0\}$ denote its {\em multiplicative group}, 
where $q=p^h$ is a prime power. 
An $[n,k,d]_q$ {\em linear code} $\C$ is a $k$-dimensional linear subspace of $\F_q^n$ with minimum distance $d:=d(\C)$. 
The parameters of $\C$ satisfy the {\em Singleton bound} $d\leq n-k+1$ \cite{LX2004}.
Given an $[n,k]_q$ or $[n,k]_{q^2}$ linear code $\C$, 
its {\em (Euclidean) dual} and {\em Hermitian dual} are given by 
\begin{align*}
  \begin{split}
    & \C^{\perp}:=\left\{\yyy=(y_1,y_2,\ldots,y_n)\in \F_q^n:~\sum_{i=1}^nx_iy_i=0,~\forall~ \xxx=(x_1,x_2,\ldots,x_n)\in \C\right\},~{\rm and} \\
    & \C^{\perp_{\rm H}}:=\left\{\yyy=(y_1,y_2,\ldots,y_n)\in \F_{q^2}^n:~\sum_{i=1}^nx_iy_i^q=0,~\forall~ \xxx=(x_1,x_2,\ldots,x_n)\in \C\right\},~{\rm respectively.}    
  \end{split}
\end{align*}
The code $\C$ is called a {\em maximum distance separable (MDS) code} if $d(\C)=n-k+1$, and a {\em near MDS (NMDS) code} if $d(\C)=n-k$ and $d(\C^{\perp})=k$. 
In addition, $\C$ is said to be a {\em self-orthogonal code} if $\C\subseteq \C^{\perp}$, and a {\em dual-containing code} if $\C^{\perp}\subseteq \C$.  
{All of these definitions also apply when the Hermitian dual is used in place of the Euclidean dual.}

\subsection{Classical locally recoverable codes}

{\em Classical locally recoverable codes (cLRCs)} were introduced to repair a single failed node 
in modern distributed storage systems (DSSs) in \cite{GHSY2012}. 
Specifically, for each $i\in \{1,2,\ldots,n\}$ and each codeword $\ccc=(c_1,c_2,\ldots,c_n)\in \C$, 
if the $i$-th symbol $c_i$ can be recovered by accessing {\em at most $r$} other code symbols of $\ccc$, 
then the $[n,k,d]_q$ linear code $\C$ is called an $(n,k,d,q;r)$-cLRC, or simply an $r$-cLRC. 
Later, the concept was further generalized to the {\em $(r, \delta)$-cLRC} by Prakash $et~al.$ \cite{PKLK2012} 
to address the situation of multiple device failures. 
Formally, the $i$-th symbol $c_i$ of an $[n,k,d]_q$ linear code $\C$ is said to have $(r,\delta)$-locality  
if there exists a subset $S_i\subseteq \{1,2,\ldots,n\}$ with $i\in S_i$ and $|S_i|\leq r+\delta-1$ 
such that the punctured code $\C|_{S_i}$ has minimum distance at least $\delta$, $i.e.,$ $d(\C|_{S_i})\geq \delta$, 
where $\C|_{S_i}$ denotes the code obtained by deleting the components indexed by the set $\{1,2,\ldots,n\}\setminus S_i$ 
in each codeword of $\C$. If all symbols of $\C$ possess $(r,\delta)$-locality, 
then $\C$ is referred to as an $(r,\delta)$-cLRC. 
In particular, $r$-cLRCs coincide with $(r,\delta)$-cLRCs for $\delta=2$. 
In the past decade, both $r$-cLRCs and $(r,\delta)$-cLRCs have been extensively studied, particularly with respect to their bounds and constructions; see \cite{FTFCX2024,ABHMT2018,CM2013,CWLX2021,CQLLXZ2024,GFWH2019,GJR2023,KNF2019,TB2014,LEL2023,P2025,HYM2025}. 
In addition, it has been practically observed that the failure of a single storage node is a more common occurrence 
than the catastrophic failure of multiple nodes simultaneously in large-scale classical DSSs, 
which further increases the interest in $r$-cLRCs \cite{SRT2025}.

\subsection{Quantum locally recoverable codes}

{With the rapid development of quantum computing and quantum storage theory} (see for example \cite{S2025,NKV2024,HEHBANS2016,MRELW2023,LMSDL2023}), 
quantum data storage could be realized in the future, which demands the necessity of quantum counterparts of cLRCs, 
namely {\em quantum locally recoverable codes (qLRCs)}. 
Very recently, Golowic and Guruswami \cite{GG-QLRC2025} introduced qLRCs with locality $r$, 
which are quantum codes $\mathcal{Q}$ with the requirement that, 
if any single qudit of $|\varphi\rangle \in \mathcal{Q}$ is erased, 
then each state $|\varphi\rangle\in \mathcal{Q}$ can be recovered by $r$ other qudits of $|\varphi\rangle$ in a recovery channel. 
They also proposed a connection between qLRCs and quantum low-density parity-check (qLDPC) codes,
highlighting that qLRCs serve as a foundational step toward studying stronger locality properties in qLDPC codes. 
In addition, for quantum codes derived from the classical CSS construction \cite{KKKS2006}, 
the said authors established a useful link between the localities of the quantum codes and those of the classical codes employed.

Based on this correspondence, Bu {\em et al.} \cite{BGL2025}, Golowic and Guruswami \cite{GG-QLRC2025}, Luo {\em et al.} \cite{LCEL2025}, 
Sharma {\em et al.} \cite{SRT2025}, and Xie {\em et al.} \cite{XZS2025} further constructed several families of qLRCs 
by employing a variation of the hypergraph product, classical Tamo-Barg codes \cite{TB2014}, 
parity-check matrices and cyclic codes, good polynomials, and trace codes, respectively. 
We summarize their parameters in Table~\ref{tab:qLRCs}, from which one can observe specific limitations with qLRCs 
derived from quantum CSS codes, particularly in terms of constrained localities, small minimum distances, and fixed code dimensions.
Recognizing these limitations of the CSS construction, Luo $et~al.$ \cite{LCEL2025} proposed the following open problem:

\begin{problem}{\em (\!\! \cite[Section V]{LCEL2025})}\label{prob.qLRCs_from_Hermitian}
Can we use methods, particularly the Hermitian construction (other than the CSS approach), to construct qLRCs via classical codes? 
\end{problem} 

Recently, $r$-qLRCs were generalized to {\em $(r,\delta)$-qLRCs} by Galindo $et~al.$ in \cite{GHMM2024}.  
Similar to the classical setup, we again abbreviate them as $r$-qLRCs, or just qLRCs for $\delta=2$, in the sequel. 
In the same work, the authors further explored the relationship between the localities of qLRCs and those of classical codes under 
the Euclidean, Hermitian, and symplectic inner products. 
By applying the Hermitian construction for quantum codes \cite{KKKS2006}, they demonstrated the existence of optimal $(r,\delta)$-qLRCs, assuming the availability of Hermitian dual-containing MDS codes. 
Such a result can also be reduced to the case of $r$-qLRCs. 
{However, since any $[n,k]_q$ MDS code inherently possesses the largest locality $k$}, the resulted qLRC may be considered 
trivial or impractical from an application-oriented perspective \cite{TB2014,GHSY2012}.

\subsection{Our Contributions}\label{sec1.3}
From the above discussions, it remains open whether there are nontrivial answers to Open Problem \ref{prob.qLRCs_from_Hermitian}. 
Furthermore, {\em can we employ the Hermitian construction to obtain qLRCs with good and more flexible parameters?} 
This paper delves into this problem. Our main contributions can be summarized as follows. 

\begin{enumerate}
  \item To measure the performance of qLRCs derived from the Hermitian construction, 
    we study bounds for the parameters of these qLRCs. 
    Specifically, we obtain the following results: 
    \begin{itemize}
      \item We establish the Griesmer-like bound, the CM-like bound, the Singleton-like bound, and the Plotkin-like bound for 
      qLRCs derived from the Hermitian construction presented in Lemma \ref{lem.Hermitian_qLRC}, 
      by exploring the connections between qLRCs and cLRCs in Theorem \ref{th.qLRCs_bounds}. 

      \item In terms of the asymptotic formulas obtained in \eqref{eq.Griesmer-like_bound_asymptotic}, \eqref{eq.CM-like_bound_asymptotic}, 
    \eqref{eq.Singleton-like_bound_asymptotic}, and \eqref{eq.Plotkin-like_bound_asymptotic}, we make detailed comparisons among the four bounds in Subsection \ref{sec.comparisons_bounds}.  
    The results show that the CM-like bound is the tightest one, while the others have simpler forms and their tightness changes with the parameters. 

    \item For different parameters, three figures are provided in Figures \ref{fig.1}, \ref{fig.2}, and \ref{fig.3} to intuitively illustrate the difference of these bounds. 
    We emphasize in Remark \ref{rem.qLRCs_bounds111} that our Singleton-like bound improves upon the one in \cite{GHMM2024}.
    \end{itemize}

  \item From the Hermitian construction in Lemma \ref{lem.Hermitian_qLRC}, we need to construct Hermitian dual-containing codes with determined localities to obtain qLRCs.
  Together with the fact that NMDS codes supporting $t$-designs yield cLRCs that are optimal with respect to the classical Singleton-like bound and the CM bound \cite{TFDTZ2023}, 
  we construct Hermitian dual-containing codes with determined localities {from the NMDS codes}, called Han-Zhang codes that introduced by Han and Zhang in \cite{HZ2024}. 
  Specifically, we get the following results: 
  \begin{itemize}
    \item Using the subset sum theory, we prove that both the minimum weight codewords of NMDS Han-Zhang codes and their duals 
    support new $t$-designs for $t\in \{2,3\}$ in Theorems \ref{th.333design} and \ref{th.222design}, respectively.
    Note that known families of linear codes supporting $t$-designs with $t\geq 2$ usually have fixed and small dimensions \cite{DT2022,DT2020,HW2023,HWL2023,XCQ2022,XCLW2024,XLX2025,DSY2024},
    while the dimensions of our NMDS codes are general or flexible. 
    Further advantages are outlined in Remark \ref{rem.designs}. 

    \item We present the weight distributions of these Han-Zhang codes in Theorems \ref{th.333design_wd} and \ref{th.222design_wd}. 
    
    \item For arbitrary NMDS codes supporting $t$-designs, we show that they always deduce exact cLRCs that are optimal simultaneously with respect to 
    all the classical Griesmer-like, CM, Singleton-like, and Plotkin-like bounds in Theorem \ref{th.optimal_LRCs_from_designs}. 
    Building on this, we further derive four classes of such optimal cLRCs in Theorems \ref{th.optimal_LRC111} and \ref{th.optimal_LRC222}. 
  \end{itemize}
  
\item By exploring the Hermitian dual-containing conditions of Han-Zhang codes, 
we finally derive three explicit families of qLRCs in Theorems \ref{th.qLRC111}, \ref{th.qLRC222}, and \ref{th.qLRC333}. 
Compared to the bounds obtained in Theorem \ref{th.qLRCs_bounds}, we confirm that 
our qLRCs are optimal with respect to some bounds.
We also list our qLRCs together with those known ones derived from the CSS construction in Table \ref{tab:qLRCs},
and can conclude that our results are new and provide more flexible parameters thanks to different localities, larger minimum distances,  
and variable dimensions in Remark \ref{rem.comparisons_qLRCs}. 
As a consequence, we provide a nontrivial and affirmative answer to Open Problem \ref{prob.qLRCs_from_Hermitian}.
\end{enumerate}

The rest of this paper is organized as follows. 
In Section \ref{sec.2}, we recall some basic definitions and results on cLRCs, qLRCs, designs, and NMDS codes. 
In Section \ref{sec.3}, we study bounds for qLRCs from the Hermitian construction and make comparisons among them. 
In Section \ref{sec.4}, we construct new qLRCs from Han-Zhang codes with general or flexible dimensions supporting $t$-designs. 
In Section \ref{sec.5}, we conclude this paper and discuss further research directions.

\section{Preliminaries}\label{sec.2}

In this section, we recall some basic definitions and results on cLRCs, qLRCs, designs, and NMDS codes.

\subsection{cLRCs and qLRCs}\label{sec.3.2}

We first recall some basic results on cLRCs. 
Given a codeword $\ccc=(c_1,c_2,\ldots,c_n)\in \C$, 
define its {\em support} as $\supp(\ccc)=\{1\leq i\leq n:~c_i\neq 0\}$. 
Then a cLRC can be mathematically defined as follows. 

\begin{definition}{\em (\!\! \cite[Definition 1]{LCEL2025})}
  An $[n,k,d]_q$ linear code $\C$ with dual $\C^{\perp}$ is said to be a {\em classical locally recoverable code (cLRC) with locality $r$} 
  if, for each $i\in \{1,2,\ldots,n\}$, 
  there exists some codeword $\overline{\ccc}=(c_1,c_2,\ldots,c_n)\in \C^{\perp}$ 
  such that $i\in \supp(\overline{\ccc})$ and $|\supp(\overline{\ccc})|\leq r+1$. 
  Moreover, we abbreviate such a code $\C$ as an {$(n,k,d,q;r)$ cLRC}, or simply an {$r$-cLRC}. 
\end{definition}

Similar to classical linear codes, cLRCs exhibit various parameter trade-offs.
Several fundamental bounds on cLRCs have been established in the literature using tools from coding theory and combinatorics.
We summarize some of these bounds below for later reference. 

\begin{lemma}\label{lem.bound_LRC}
	Let $\C$ be an $(n,k,d,q;r)$ cLRC and let $k^{(q)}_{opt}(n,d)$ be the largest possible dimension of a $q$-ary linear code satisfying $1\leq d\leq n$. 
    Suppose that $\ell$ is an integer.
    Then the following bounds hold. 
    \begin{itemize}
        \item [\rm 1)] {\em (\!\! Griesmer-like bound \cite{HXSCFY2020})} {We have that}  $$n\geq \max_{1\leq \ell\leq \left\lceil \frac{k}{r} \right\rceil-1}
                                                                \left\{\ell(r+1)+\sum_{i=0}^{k-\ell r-1}\left\lceil \frac{d}{q^i} \right\rceil \right\}.$$

        \item [\rm 2)] {\em (\!\! CM bound \cite{CM2013})} We have that $$k\leq \min_{0\leq \ell\leq {\left\lfloor \frac{n-1}{r+1} \right\rfloor}}\{\ell r+k^{(q)}_{opt}(n-\ell(r+1),d)\}.$$ 

        \item [\rm 3)] {\em (\!\! Singleton-like bound \cite{GHSY2012})} We have that $$d\leq n-k-\left\lceil \frac{k}{r} \right\rceil+2.$$

        \item [\rm 4)] {\em (\!\! Plotkin-like bound \cite{HXSCFY2020})} We have that $$d\leq \min_{1\leq \ell\leq \left\lceil \frac{k}{r} \right\rceil-1}
                                                                \left\{\frac{q^{k-\ell r-1}(q-1)(n-\ell(r+1))}{q^{k-\ell r}-1} \right\}.$$
    \end{itemize}
\end{lemma}

Now, we focus on qLRCs. 
{To this end}, we keep the notation used in \cite{GHMM2024,KKKS2006,GG-QLRC2025,BGL2025,NC2011}. 
Let $\mathbb{C}$ be the {\em field of complex numbers} and let $\mathbb{C}^q$ be the $q$-dimensional {\em Hilbert space} over $\mathbb{C}$. 
Specifically, a quantum code $\mathcal{Q}$ of length $n$, dimension $\kappa$, minimum distance $\delta$, and alphabet size $q$, 
denoted by $[[n,\kappa,\delta]]_q$, is a $q^{\kappa}$-dimensional subspace of the Hilbert space 
$\mathbb{C}^q\otimes \mathbb{C}^q\otimes\ldots\otimes \mathbb{C}^q:=(\mathbb{C}^q)^{\otimes n}$. 
Such a quantum code $\mathcal{Q}$ can encode $\kappa$ logical qudits into entangled states of $n$ qudits 
and protect against the erasure of any set of $\delta - 1$ qudits. 
There are various types of quantum codes, and the so-called {\em quantum stabilizer codes} form the most-known class. 
Quantum stabilizer codes have close connections to classical codes, 
containing those derived from the CSS construction, Hermitian construction, and symplectic construction \cite{KKKS2006}. 
For brevity, we omit details about quantum (stabilizer) codes, and instead we give straightforward results for qLRCs, 
which are sufficient for understanding the later sections. 
For new notation involved, commonly used in quantum mechanics, 
please refer to \cite{GG-QLRC2025,KKKS2006,GHMM2024,NC2011,BGL2025} for further details.

\begin{definition}{\em (\!\! \cite[Definitions 8 and 9]{GHMM2024})}\label{def.qLRCs}
  A quantum code $\mathcal{Q}\subseteq (\mathbb{C}^q)^{\otimes n}$ is said to be a 
  {\em quantum locally recoverable code (qLRC) with locality $r$}, or simply an $r$-qLRC, if, 
  for each $i\in \{1,2,\ldots,n\}$, there exists a set $J\subseteq \{1,2,\ldots,n\}$ containing $i$ with $|J|\leq r+1$ 
  such that for every subset $I\subsetneq J$ with $|I|=1$, there exists a trace-preserving map $\mathcal{R}_{\mathcal{Q},I}^J$, 
  which acts only on the qudits corresponding to $J$ and keeps untouched the remaining ones, such that 
  $$
  \mathcal{R}_{\mathcal{Q},I}^J \circ \Gamma^I(|\varphi\rangle \langle \varphi|) = |\varphi\rangle \langle \varphi|
  $$
  for any $|\varphi\rangle \in \mathcal{Q}$, where $\Gamma^I$ is a mapping given as in \cite[Page 6]{GHMM2024}.
\end{definition}

For $\mathcal{Q}$ a stabilizer code, Definition \ref{def.qLRCs} can be reduced to an easier form by considering cLRCs with special structures. 
Note that the characterization of qLRCs from quantum CSS codes has been derived in \cite{GG-QLRC2025}. 
In the following, we present the Hermitian construction for qLRCs. 
To this end, we need the Hermitian construction for quantum stabilizer codes \cite{KKKS2006}. 

\begin{lemma}{\em (\!\! Hermitian construction \cite{KKKS2006})}\label{lem.Hermitian_construction} 
    If $\C$ is an $[n,k,d]_{q^2}$ Hermitian dual-containing code, 
then there exists an $[[n,\kappa,\delta]]_q$ quantum stabilizer code $\mathcal{Q}$ with 
$\kappa = 2k-n$ and $\delta = \wt(\C\setminus \C^{\perp_{\rm H}})$.  
Furthermore, the quantum stabilizer code $\mathcal{Q}$ is said to be {\em pure} if $\delta = \wt(\C)=d$;  
and {\em impure} otherwise, 
{where $\wt(\SSS)=\min\{\wt({\bf s}):~{\bf s}\in \SSS,~{\bf s}\neq {\bf 0}\}$ for any nonempty subset $\SSS\subseteq \F_{{q^2}}^n$ and 
$\wt({\bf s})$ is the Hamming weight of ${\bf s}\in \SSS$.}
\end{lemma}

Combining with Lemma \ref{lem.Hermitian_construction} and \cite[Theorem 28]{GHMM2024}, we immediately derive the Hermitian construction for qLRCs as follows. 

\begin{lemma}\label{lem.Hermitian_qLRC}
{\em (\!\! Hermitian construction for qLRCs)} 
  If $\C$ is an $[n,k,d]_{q^2}$ Hermitian dual-containing code with locality $r$ and $d(\C^{\perp_{\rm H}})\geq 2$, 
then there exists an $$[[n,\kappa,\delta]]_q$$ qLRC $\mathcal{Q}$ with locality $r$, 
where 
\begin{align*}
  \kappa  = 2k-n,~{\rm and}~\delta  = \wt(\C\setminus \C^{\perp_{\rm H}})\geq d.  
\end{align*}
Furthermore, the qLRC $\mathcal{Q}$ is said to be {\em pure} if $\delta = \wt(\C)=d$;  
and {\em impure} otherwise.
\end{lemma}

\subsection{NMDS codes supporting $t$-designs}\label{sec.3.1}

Let $1\leq t\leq k\leq n$ be three positive integers. 
Let $\mathcal{P}$ be a set with $|\mathcal{P}|=n$ and let $\mathcal{B}$ be a collection of $k$-subsets of $\mathcal{P}$. 
We call the pair $(\mathcal{P},\mathcal{B})$ a {\em{$t$-$(n,k,\lambda)$ design}} with 
$b=\frac{\lambda\binom{n}{t}}{\binom{k}{t}}$ blocks, 
if each $t$-subset of $\mathcal{P}$ is contained in exactly $\lambda$ elements of $\mathcal{B}$.  
Denote by $\mathcal{B}^{\perp}$ the set of the complements of  {all} the blocks in $\mathcal{B}$. 
If $(\mathcal{P},\mathcal{B})$ forms a $t$-$(n,k,\lambda)$ design, then $(\mathcal{P},\mathcal{B}^{\perp})$ is a 
$t$-$(n,n-k,\lambda^{\perp})$ design, where 
\begin{align}\label{eq.complementary_design}
  \lambda^{\perp}=\frac{\lambda\binom{n-t}{k}}{\binom{n-t}{k-t}}
\end{align}
and we call it the {\em{complementary design}} of $(\mathcal{P},\mathcal{B})$. 

Many linear codes induce $t$-designs \cite{DT2022}, and they are referred to as {\em linear codes supporting $t$-designs} in this paper. 
Let $\C$ be a linear code of length $n$ and $\mathcal{P}(\C)=\{1,2,\ldots,n\}$. 
Let $\mathcal{B}_w(\C)=\frac{S}{q-1}$, where  {$S$ is the multiset} given by 
$$S=\{\{\supp(\ccc):~ \wt(\ccc)=w~{\rm and}~\ccc\in \C\}\}.$$  
{Thus,} $\frac{S}{q-1}$ is the multiset derived from dividing the multiplicity of each element in $S$ by $q-1$.  
For $0\leq i\leq n$, let $A_i$ (resp. $A^{\perp}_i$) be the {\em number of codewords}  {of} weight $i$ in $\C$ (resp. $\C^{\perp}$). 
We have the following definition.  
\begin{definition}{\em (\!\!\cite{DT2022})}\label{def.t-design} 
If the pair $(\mathcal{P}(\C),  \mathcal{B}_w(\C))$ is a $t$-$(n,w,\lambda)$ design with $b$ blocks for some $0\leq w\leq n$, 
we say that {\em{the code $\C$ supports a $t$-design}}, where 
\begin{align}\label{eq.parameters of t-design}
	\lambda=\frac{A_w\binom{w}{t}}{(q-1)\binom{n}{t}},~{\rm and}~ b=\frac{A_w}{q-1}. 
\end{align}
\end{definition}

For any linear code $\C$ supporting a $t$-$(n,k,\lambda_t)$ design, it naturally supports $s$-$(n,k,\lambda_s)$ designs with 
$\lambda_s=\lambda_t \frac{\binom{n-s}{t-s}}{\binom{k-s}{t-s}}$ for any $s\leq t$ (see also \cite[Theorem 4.4]{DT2022}). 
The following result establishes a further connection between NMDS codes and $t$-designs. 

\begin{lemma}{\em (\!\! \cite[Proposition 14]{FW1997-sAMDS})}\label{lem.complement design}
	Let $\C$ be an $[n,k,n-k]_q$ NMDS code. 
	For any minimum weight codeword $\ccc$ in $\C$, there exists, up to a multiple, a unique minimum weight codeword $\ccc^{\perp}$ in $\C^{\perp}$ 
	satisfying 
  $$\supp(\ccc)\cap \supp(\ccc^{\perp})=\emptyset.$$ 
	Moreover, the number of minimum weight codewords in $\C$ and the number of those in $\C^{\perp}$ are equal, 
  $i.e., A_{n-k}=A^{\perp}_{k}$. 
\end{lemma}

Combining Lemma \ref{lem.complement design} with \eqref{eq.complementary_design} and \eqref{eq.parameters of t-design}, 
if the minimum weight codewords of an $[n,k,n-k]_q$ NMDS code support a $t$-$(n,n-k,\lambda)$ design, 
then the minimum weight codewords of its dual support a $t$-$(n,k,\lambda^{\perp})$ design,  
where 
\begin{align}\label{eq.complementary_design_parameters}
	\lambda^{\perp}=\frac{\lambda\binom{n-t}{n-k}}{\binom{n-t}{n-k-t}}.    
\end{align}

Let $\{A_i:~ i=0,1,\ldots,n\}$ (resp. $\{A^{\perp}_i:~ i=0,1,\ldots,n\}$) denote the {\em weight distribution} of $\C$ (resp. $\C^{\perp}$). 
Furthermore, we use $A(z)=1+A_1z+A_2z^2+\ldots+A_nz^{n}$ and $A^{\perp}(z)=1+A^{\perp}_1z+A^{\perp}_2z^2+\ldots+A^{\perp}_nz^{n}$ to denote 
the {\em polynomial weight enumerators} of $\C$ and $\C^{\perp}$, respectively. 
In \cite{DL1995-NMDS}, Dodunekov and Landgev proved that weight distributions of an NMDS code and its dual are uniquely 
determined by the numbers of their minimum weight codewords, respectively.   

\begin{lemma}{\em (\!\! \cite[Corollary 4.2]{DL1995-NMDS})}\label{lem.NMDS weight distribution}
	Let $\C$ be an $[n,k,n-k]_q$ NMDS code. If $s\in \{1,2,\ldots,k\}$, then 
	$$A_{n-k+s}=\binom{n}{k-s}\sum_{i=0}^{s-1}(-1)^i\binom{n-k+s}{i}(q^{s-i}-1)+(-1)^s\binom{k}{s}A_{n-k}.$$ 
	If $s\in \{1,2,\ldots,n-k\}$, then 
	$$A^{\perp}_{k+s}=\binom{n}{k+s}\sum_{i=0}^{s-1}(-1)^i\binom{k+s}{i}(q^{s-i}-1)+(-1)^s\binom{n-k}{s}A^{\perp}_{k}.$$ 
\end{lemma}

\section{Bounds and comparisons for qLRCs from the Hermitian construction}\label{sec.3}

In this section, we present several bounds for qLRCs derived from the Hermitian construction  
and make detailed comparisons among them. 
{These bounds are} necessary for measuring the performance of a qLRC. 

\subsection{Several bounds for qLRCs from the Hermitian construction}

Luo $et~al.$ proposed a Singleton-like bound and a CM-like bound for qLRCs from the CSS construction in \cite[Corollaries 5 and 6]{LCEL2025}. 
In \cite[Theorem 30]{GHMM2024}, Galindo $et~al.$ presented a Singleton-like bound for qLRCs 
from the Hermitian construction described in Lemma \ref{lem.Hermitian_qLRC}. However, this bound applies only to pure qLRCs (see Remark \ref{rem.qLRCs_bounds111}). 
Moreover, to the best of our knowledge, no more results have been reported on other bounds for qLRCs obtained via the Hermitian construction. 
The following result provides an improved Singleton-like bound and three new bounds for qLRCs from the Hermitian construction.

\begin{theorem}\label{th.qLRCs_bounds}
    Let $\mathcal{Q}$ be an $[[n,\kappa,\delta]]_q$ qLRC with locality $r$ constructed by the Hermitian construction given in Lemma \ref{lem.Hermitian_qLRC}.
  Then the following bounds hold. 
  \begin{itemize}
    \item [\rm 1)] {\em (Griesmer-like bound)} 
    If $r< \kappa$, we have that 
  \begin{align}\label{eq.Griesmer-like_bound}
    n+\kappa \geq 2\cdot \max_{1\leq \ell\leq \left\lceil \frac{\kappa}{r} \right\rceil-1}
                                                                \left\{\ell(r+1)+\sum_{i=0}^{\kappa-\ell r-1}\left\lceil \frac{\delta}{q^{2i}} \right\rceil \right\}. 
  \end{align}

      \item [\rm 2)] {\em (CM-like bound)} 
 We have that 
  \begin{align}\label{eq.CM-like_bound}
      \kappa \leq \min_{0\leq \ell\leq \left\lceil \frac{n+\kappa-1}{2(r+1)} \right\rceil-1}\left\{\ell r+k_{opt}^{(q^2)}\left(\frac{n+\kappa}{2}-\ell(r+1),\delta\right)\right\}.
  \end{align}

      \item [\rm 3)] {\em (Singleton-like bound)} 
    We have that 
  \begin{align}\label{eq.Singleton-like_bound}
    2\delta \leq n-\kappa-2\left\lceil \frac{\kappa}{r} \right\rceil+4. 
  \end{align}

      \item [\rm 4)] {\em (Plotkin-like bound)}
    If $r< \kappa$, we have that 
  \begin{align}\label{eq.Plotkin-like_bound}
    2\delta\leq \min_{1\leq \ell\leq \left\lceil \frac{\kappa}{r} \right\rceil-1}
                                                                \left\{\frac{q^{2\kappa-2\ell r-2}(q^2-1)(n+\kappa-2\ell(r+1))}{q^{2\kappa-2\ell r}-1} \right\}. 
  \end{align}

  \end{itemize} 
\end{theorem}
\begin{IEEEproof}
  According to the assumption, it follows from Lemma \ref{lem.Hermitian_qLRC} that 
  there exists an $[n,\frac{n+\kappa}{2},d\leq \delta]_{q^2}$ Hermitian dual-containing code $\C$ with $d(\C^{\perp_{\rm H}})\geq 2$ and locality $r$. 
  Since $\C^{\perp_{\rm H}}\subseteq \C$, we can assume that $\C$ has a generator matrix of the form 
  \begin{align*}
    \left(\begin{array}{cc}
      I_{\frac{n-\kappa}{2}} & G \\
      O_{\kappa\times \frac{n-\kappa}{2}} & P
    \end{array}\right),
  \end{align*}
  where $(I_{\frac{n-\kappa}{2}}~ G)$ generates $\C^{\perp_{\rm H}}$. 
  It then implies that $(O_{\kappa\times \frac{n-\kappa}{2}}~ P)$ can generate an $[n,\kappa,d']_{q^2}$ linear code $\C'$ 
  and $d'\geq \wt(\C\setminus \C^{\perp_{\rm H}})=\delta$, as any nonzero codeword of $\C'$ always belongs to $\C\setminus \C^{\perp_{\rm H}}$. 
  Combining with the fact that $\C'\subseteq \C$, we immediately deduce that locality of $\C'$ is at most $r$. 
  By deleting the first $\frac{n-\kappa}{2}$ coordinates of $\C'$, we then obtain an $[\frac{n+\kappa}{2},\kappa,d''\geq \delta]_{q^2}$ linear code $\C''$ 
  ($i.e.,$ the linear code generated by $P$)  with locality at most $r$. 
  {Therefore, the cLRC $\C''$ has parameters}
  \begin{align}\label{eq.C''}
    \left(\frac{n+\kappa}{2},\kappa,d''\geq \delta,q^2;r''\leq r\right).
  \end{align}

  Let 
\begin{align}\label{eq.opt_nkd}
  \begin{split}
  & n_{opt}^{(q^2)}(k^*,d^*;r^*):=\min\{n:~{\rm there~is~an~} (n^*,k^*,d^*,q^2;r^*) ~{\rm cLRC}\},\\
  & k_{opt}^{(q^2)}(n^*,d^*;r^*):=\max\{k:~{\rm there~is~an~} (n^*,k^*,d^*,q^2;r^*) ~{\rm cLRC}\},\\
  & d_{opt}^{(q^2)}(n^*,k^*;r^*):=\max\{d:~{\rm there~is~an~} (n^*,k^*,d^*,q^2;r^*) ~{\rm cLRC}\}.
  \end{split}
\end{align}
It is evident that if a symbol in a linear code is recoverable from accessing at most $r$ other symbols, 
then it is also recoverable from accessing any superset of those symbols, including $r+1$ symbols.
Applying \eqref{eq.opt_nkd} to the code $\C''$ with the parameters given in \eqref{eq.C''}, 
it can be checked that  
\begin{align}
  & n+\kappa\geq  2n_{opt}^{(q^2)}(\kappa,d'';r'')\geq 2n_{opt}^{(q^2)}(\kappa,\delta;r), \label{eq.qLRC_n} \\ 
  & \kappa\leq k_{opt}^{(q^2)}\left(\frac{n+\kappa}{2},d'';r''\right)\leq k_{opt}^{(q^2)}\left(\frac{n+\kappa}{2},\delta;r\right), \label{eq.qLRC_k}\\
  & \delta\leq d''\leq d_{opt}^{(q^2)}\left(\frac{n+\kappa}{2},\kappa;r''\right)\leq d_{opt}^{(q^2)}\left(\frac{n+\kappa}{2},\kappa;r\right). \label{eq.qLRC_d}
\end{align}

Take the Plotkin-like bound as an example. 
Applying the Plotkin-like bound for cLRCs presented in Lemma \ref{lem.bound_LRC}.4) to $\C''$, 
it follows from the assumption $r<\kappa$ and \eqref{eq.qLRC_d} that 
\begin{align*}
2\delta\leq \min_{1\leq \ell\leq \left\lceil \frac{\kappa}{r} \right\rceil-1}
                                                                \left\{\frac{q^{2\kappa-2\ell r-2}(q^2-1)(n+\kappa-2\ell(r+1))}{q^{2\kappa-2\ell r}-1} \right\}.
\end{align*}
{The other three bounds can be similarly derived from applying Lemmas \ref{lem.bound_LRC}.1)-3) to \eqref{eq.qLRC_n}-\eqref{eq.qLRC_d}, respectively, 
noting that
$$\left\lfloor \frac{\frac{n+\kappa}{2}-1}{r+1} \right\rfloor=\left\lfloor \frac{n+k-2}{2(r+1)} \right\rfloor=\left\lceil \frac{n+k-1}{2(r+1)} \right\rceil-1.$$ }
This completes the proof.
\end{IEEEproof}

\begin{remark}\label{rem.optimal_qLRCs}
  We say that an $[[n,\kappa,\delta]]_q$ qLRC $\mathcal{Q}$ with locality $r$ is {\em optimal} with respect to a certain bound, 
  if its parameters make the the equality hold in that bound. 
\end{remark}

\begin{remark}\label{rem.qLRCs_bounds111}
  We give some remarks on our bounds in Theorem \ref{th.qLRCs_bounds}. 
  \begin{enumerate}
    \item Note that bounds given in Theorem \ref{th.qLRCs_bounds} are available for both {\em pure} and {\em impure} qLRCs. 
Particularly, for {\em pure} $[[n,\kappa,\delta]]_q$ qLRCs obtained from the Hermitian construction, 
applying a similar line of reasoning as in the proof of Theorem \ref{th.qLRCs_bounds} to the corresponding 
$[n,\frac{n+\kappa}{2},\delta]_{q^2}$ Hermitian dual-containing code $\C$ with $d(\C^{\perp_{\rm H}})\geq 2$ and locality $r$ 
immediately gives the {\em pure Singleton-like bound} as follows: 
  \begin{align}\label{eq.Singleton-like_bound_pure}
    2\delta \leq n-\kappa-2\left\lceil \frac{n+\kappa}{2r} \right\rceil+4,
  \end{align}
  which coincides with the bound proposed in \cite[Theorem 30]{GHMM2024}. 
  However, if $\mathcal{Q}$ is impure, the bound in (\ref{eq.Singleton-like_bound_pure}) becomes inapplicable 
  due to the lack of information about the minimum distance of $\mathcal{Q}$, 
  whereas our bound in (\ref{eq.Singleton-like_bound}) remains valid.
  As a consequence, Theorem \ref{th.qLRCs_bounds}.3) is an improvement of \cite[Theorem 30]{GHMM2024} for impure cases. 

    \item   
Note also that \eqref{eq.qLRC_n}-\eqref{eq.qLRC_d} actually provide more general bounds for qLRCs from the Hermitian construction, 
and they can be further specialized to useful bounds by combining with some known bounds for cLRCs, 
for example, those proposed in \cite{ABHMT2018,GFWH2019,KNF2019,CQLLXZ2024,FTFCX2024} and the references therein. 
We omit these details here.
  \end{enumerate}
\end{remark}

\subsection{Comparisons of these bounds}\label{sec.comparisons_bounds}

To compare the bounds obtained in Theorem \ref{th.qLRCs_bounds}, it is better to consider their asymptotic version for $n\to \infty$. 
Given an $[[n,\kappa,\delta]]_q$ qLRC with locality $r$, we let 
\begin{align*}
  \mathcal{R}=\frac{\kappa}{n},~{\rm and}~\Delta=\frac{\delta}{n}
\end{align*}
denote its {\em rate} and {\em relative distance}, respectively. 
In the following, we fix the locality $r$ and the field size $q$. 

Let $t$ and $\ell_0$ be two integers satisfying $1\leq t\leq \kappa-\ell_0 r$ and $1\leq \ell_0\leq \left\lceil \frac{\kappa}{r}\right\rceil-1$. 
From the Griesmer-like bound in \eqref{eq.Griesmer-like_bound}, 
we have 
\begin{align*}
  \begin{split}
    n+\kappa  \geq 2\cdot \left(\ell_0(r+1)+\sum_{i=0}^{\kappa-\ell_0 r-1} \left\lceil \frac{\delta}{q^{2i}} \right\rceil\right)  
     \geq 2\cdot \left(\ell_0(r+1)+\sum_{i=0}^{t-1} \frac{\delta}{q^{2i}}+\kappa-\ell_0 r-t \right) 
     = 2\kappa +\frac{2(q^{2t}-1)}{q^{2t}-q^{2t-2}}\cdot \delta+2\ell_0-2t, 
  \end{split}
\end{align*}
which implies that 
\begin{align}\label{eq.Griesmer-like_bound_asymptotic}
  {\mathcal{R}}\leq 1-\frac{2(q^{2t}-1)}{q^{2t}-q^{2t-2}}\cdot \Delta+o(1),~n\to \infty.
\end{align}
Note that, if $r\leq \left\lfloor\frac{q^2}{2(q^2-1)}(\delta-1)-\frac{3}{4} \right\rfloor$, then 
$\left\lceil \frac{1}{r+1}\left(\frac{n+\kappa}{2}-\frac{q^2}{q^2-1}(\delta-1)\right)\right\rceil
\leq \left\lceil \frac{n+\kappa-1}{2(r+1)}\right\rceil-1$. 
Taking $\ell_0=\left\lceil \frac{1}{r+1}\left(\frac{n+\kappa}{2}-\frac{q^2}{q^2-1}(\delta-1)\right)\right\rceil$ 
and a similar argument as that presented in \cite[Page 1797]{LCEL2025}, 
we conclude from the CM-like bound in \eqref{eq.CM-like_bound} and the Singleton-like bound in \eqref{eq.Singleton-like_bound} that  
\begin{align}\label{eq.CM-like_bound_asymptotic}
   {\mathcal{R}} \leq \frac{r}{r+2}-\frac{2r}{r+2}\cdot \frac{q^2}{q^2-1}\cdot \Delta+o(1),~n\to \infty,
\end{align}
and 
\begin{align}\label{eq.Singleton-like_bound_asymptotic}
  {\mathcal{R}} \leq \frac{r}{r+2}-\frac{2r}{r+2}\cdot \Delta+o(1),~n\to \infty,
\end{align}
respectively. 
Particularly, if $r\mid \kappa$ and $r\neq \kappa$, then taking $\ell=\left\lceil \frac{\kappa}{r}\right\rceil-1=\frac{\kappa}{r}-1$ 
in the Plotkin-like bound in (\ref{eq.Plotkin-like_bound}) gives that 
\begin{align*}
  2\delta \leq \frac{q^{2r}-q^{2r-2}}{q^{2r}-1}\cdot \left(n-\frac{r+2}{r}\cdot \kappa +2r+2\right),
\end{align*}
which implies that 
\begin{align}\label{eq.Plotkin-like_bound_asymptotic}
  {\mathcal{R}} \leq \frac{r}{r+2}-\frac{2r}{r+2}\cdot \frac{q^{2r}-1}{q^{2r}-q^{2r-2}}\cdot \Delta+o(1),~n\to \infty.
\end{align}

In terms of the above asymptotic formulas in \eqref{eq.Griesmer-like_bound_asymptotic}-\eqref{eq.Plotkin-like_bound_asymptotic}, we make detailed comparisons among the four bounds. 
First of all, since 
\begin{align*}
  \frac{q^2}{q^2-1}>\frac{q^{2r}-1}{q^{2r}-q^{2r-2}}>1, 
\end{align*}
the CM-like bound  in \eqref{eq.CM-like_bound_asymptotic} is tighter than the Plotkin-like bound  in \eqref{eq.Plotkin-like_bound_asymptotic}, 
which is in turn tighter than the Singleton-like bound  in \eqref{eq.Singleton-like_bound_asymptotic}. 
In addition, one can quickly check that the following results hold:  
\begin{itemize}
\item {\em (\!\! The CM-like bound in \eqref{eq.CM-like_bound_asymptotic} vs the Griesmer-like bound  in \eqref{eq.Griesmer-like_bound_asymptotic})}  
  {If $r\geq 2q^{2t}-2$,} then the CM-like bound in \eqref{eq.CM-like_bound_asymptotic} is always tighter than 
the Griesmer-like bound  in \eqref{eq.Griesmer-like_bound_asymptotic}; 
and if $r<2q^{2t}-2$, 
then the CM-like bound  in \eqref{eq.CM-like_bound_asymptotic} is tighter than 
the Griesmer-like bound  in \eqref{eq.Griesmer-like_bound_asymptotic} if and only if 
\begin{align*}
  \Delta\leq \frac{(q^2-1)q^{2t-2}}{2q^{2t}-r-2}, 
\end{align*}
and {this inequality always holds due to the fact that $\mathcal{R}\geq 0$}. 
As a result, the CM-like bound  in \eqref{eq.CM-like_bound_asymptotic} is always tighter than the Griesmer-like bound  in \eqref{eq.Griesmer-like_bound_asymptotic}.

\item {\em (\!\! The Singleton-like bound  in \eqref{eq.Singleton-like_bound_asymptotic} vs the Griesmer-like bound  in \eqref{eq.Griesmer-like_bound_asymptotic})} 
Since $\frac{q^{2t}-1}{q^{2t}-q^{2t-2}}>\frac{r}{r+2}\cdot \frac{q^{2r}-1}{q^{2r}-q^{2r-2}}$,  
the Griesmer-like bound  in \eqref{eq.Griesmer-like_bound_asymptotic} is tighter than the Singleton-like bound  in \eqref{eq.Singleton-like_bound_asymptotic} 
if and only if 
\begin{align*}
  \Delta\geq \frac{(q^2-1)q^{2t}}{2q^{2t+2}+rq^{2t}-(r+2)q^2}. 
\end{align*}

  \item {\em (\!\! The Plotkin-like bound  in \eqref{eq.Plotkin-like_bound_asymptotic} vs the Griesmer-like bound  in \eqref{eq.Griesmer-like_bound_asymptotic})}  
  {If $\frac{q^{2t}-1}{q^{2t}-q^{2t-2}}\leq \frac{r}{r+2}\cdot \frac{q^{2r}-1}{q^{2r}-q^{2r-2}}$,} 
then the Plotkin-like bound  in \eqref{eq.Plotkin-like_bound_asymptotic} is always tighter than 
the Griesmer-like bound  in \eqref{eq.Griesmer-like_bound_asymptotic}; 
and if $\frac{q^{2t}-1}{q^{2t}-q^{2t-2}}>\frac{r}{r+2}\cdot \frac{q^{2r}-1}{q^{2r}-q^{2r-2}}$, 
then the Plotkin-like bound  in \eqref{eq.Plotkin-like_bound_asymptotic} is tighter than 
the Griesmer-like bound  in \eqref{eq.Griesmer-like_bound_asymptotic} if and only if 
\begin{align*}
  \Delta\leq \frac{(q^{2}-1)q^{2r+2t}}{(q^{2t}-1)((r+2)q^{2r+2}-rq^{2t+2})}.
\end{align*}
\end{itemize}

To make the comparisons more intuitive, we present the four bounds in Figures \ref{fig.1}-\ref{fig.3} 
for $q$-ary qLRCs with locality $r$, for different values of $t$.

\begin{remark}
  From the above comparisons, we conclude that the CM-like bound in \eqref{eq.CM-like_bound_asymptotic} is the tightest one among the four bounds. 
  {However, the exact value of $k_{opt}^{(q^2)}\left(\frac{n+\kappa}{2},\delta\right)$ contained within the bound is generally difficult to determine.} 
  Instead, the other three bounds have more explicit forms and their values can be easily determined for given qLRCs. 
  Additionally, as can be seen in Figures \ref{fig.2} and \ref{fig.3}, the CM-like bound in \eqref{eq.CM-like_bound_asymptotic} 
  and the Plotkin-like bound in \eqref{eq.Plotkin-like_bound_asymptotic} can be very close to each other. 
\end{remark}

\begin{figure}[htbp]
  \caption{Comparisons of the asymptotic bounds on {qubit codes} with locality 2 for $t\in \{1,2,3\}$}\label{fig.1}
\centering
\begin{tikzpicture}
\begin{axis}[
    width=10cm, height=7cm,
    xlabel={$ \Delta= \frac{\delta}{n}$},
    ylabel={$ \mathcal{R}=\frac{\kappa}{n} $},
    xmin=0, xmax=0.6,
    ymin=0, ymax=1.1,
    legend pos=north east,  
    grid=both,
    major grid style={line width=.2pt,draw=gray!50},
    minor tick num=1,
    thick,
    enlargelimits=false,
    legend style={
    font=\footnotesize,
    at={(1.02,1)},
    anchor=north west,
    align=left,
    draw=none,
    cells={anchor=west}}
]

\pgfmathsetmacro{\q}{2}
\pgfmathsetmacro{\r}{2}
\pgfmathsetmacro{\t}{1}
\pgfmathsetmacro{\tt}{2}
\pgfmathsetmacro{\ttt}{3}
\addplot[
    red,
    thick
] expression[
    domain=0:0.5,
    samples=200
]{
    1 - 2*((\q^(2*\t) - 1)/(\q^(2*\t) - \q^(2*\t - 2)))*x
};
\addlegendentry{Griesmer-like bound in \eqref{eq.Griesmer-like_bound_asymptotic} for $t=1$}

\addplot[
    yellow!60!black,
    thick
] expression[
    domain=0:0.5,
    samples=200
]{
    1 - 2*((\q^(2*\tt) - 1)/(\q^(2*\tt) - \q^(2*\tt - 2)))*x
};
\addlegendentry{Griesmer-like bound in \eqref{eq.Griesmer-like_bound_asymptotic} for $t=2$}

\addplot[
    green,
    thick
] expression[
    domain=0:0.5,
    samples=200
]{
    1 - 2*((\q^(2*\ttt) - 1)/(\q^(2*\ttt) - \q^(2*\ttt - 2)))*x
};
\addlegendentry{Griesmer-like bound in \eqref{eq.Griesmer-like_bound_asymptotic} for $t=3$}

\addplot[
    blue,
    thick,
    densely dashed
] expression[
    domain=0:0.5,
    samples=200
]{
    (\r/(\r + 2)) - 2*(\r/(\r + 2))*(\q^2 / (\q^2 - 1))*x
};
\addlegendentry{CM-like bound in \eqref{eq.CM-like_bound_asymptotic}}

\addplot[
    green!60!black,
    thick,
    dotted
] expression[
    domain=0:0.5,
    samples=200
]{
    (\r/(\r + 2)) - 2*(\r/(\r + 2))*x
};
\addlegendentry{Singleton-like bound in \eqref{eq.Singleton-like_bound_asymptotic}}

\addplot[
    purple,
    thick,
    dashdotted
] expression[
    domain=0:0.5,
    samples=200
]{
    (\r/(\r + 2)) - 2*(\r/(\r + 2))*((\q^(2*\r) - 1)/(\q^(2*\r) - \q^(2*\r - 2)))*x
};
\addlegendentry{Plotkin-like bound in \eqref{eq.Plotkin-like_bound_asymptotic}}

\end{axis}
\end{tikzpicture}
\end{figure}
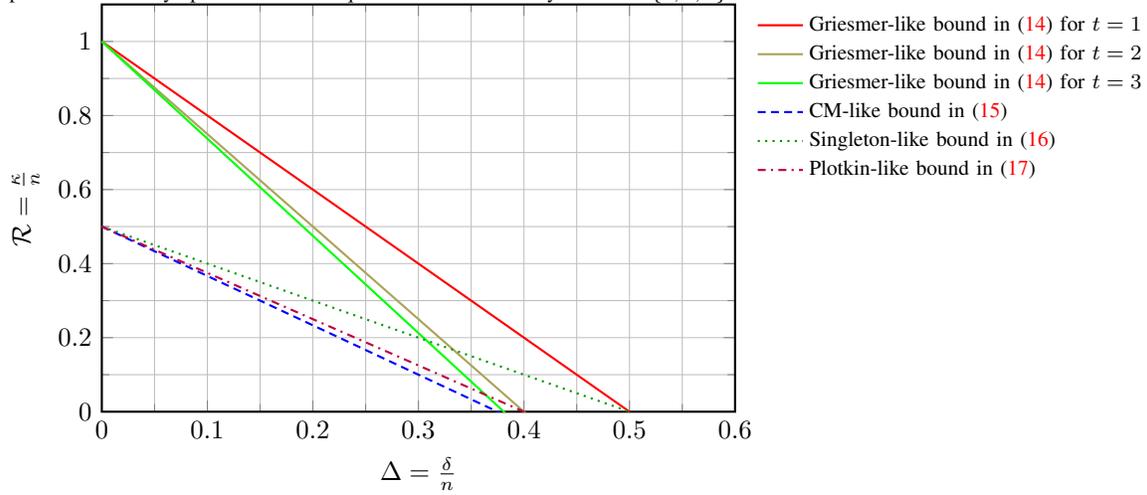

\begin{figure}[htbp]
  \caption{Comparisons of the asymptotic bounds on {qubit codes} with locality 6 for $t\in \{1,2,3\}$}\label{fig.2}
\centering
\begin{tikzpicture}
\begin{axis}[
    width=10cm, height=7cm,
    xlabel={$ \Delta= \frac{\delta}{n}$},
    ylabel={$ \mathcal{R}=\frac{\kappa}{n} $},
    xmin=0, xmax=0.6,
    ymin=0, ymax=1.1,
    legend pos=north east,  
    grid=both,
    major grid style={line width=.2pt,draw=gray!50},
    minor tick num=1,
    thick,
    enlargelimits=false,
legend style={
    font=\footnotesize,
    at={(1.02,1)},
    anchor=north west,
    align=left,
    draw=none,
    cells={anchor=west}}
]

\pgfmathsetmacro{\q}{2}
\pgfmathsetmacro{\r}{6}
\pgfmathsetmacro{\t}{1}
\pgfmathsetmacro{\tt}{2}
\pgfmathsetmacro{\ttt}{3}
\addplot[
    red,
    thick
] expression[
    domain=0:0.5,
    samples=200
]{
    1 - 2*((\q^(2*\t) - 1)/(\q^(2*\t) - \q^(2*\t - 2)))*x
};
\addlegendentry{Griesmer-like bound in \eqref{eq.Griesmer-like_bound_asymptotic} for $t=1$}

\addplot[
    yellow!60!black,
    thick
] expression[
    domain=0:0.5,
    samples=200
]{
    1 - 2*((\q^(2*\tt) - 1)/(\q^(2*\tt) - \q^(2*\tt - 2)))*x
};
\addlegendentry{Griesmer-like bound in \eqref{eq.Griesmer-like_bound_asymptotic} for $t=2$}

\addplot[
    green,
    thick
] expression[
    domain=0:0.5,
    samples=200
]{
    1 - 2*((\q^(2*\ttt) - 1)/(\q^(2*\ttt) - \q^(2*\ttt - 2)))*x
};
\addlegendentry{Griesmer-like bound in \eqref{eq.Griesmer-like_bound_asymptotic} for $t=3$}

\addplot[
    blue,
    thick,
  densely dashed,
] expression[
    domain=0:0.5,
    samples=200
]{
    (\r/(\r + 2)) - 2*(\r/(\r + 2))*(\q^2 / (\q^2 - 1))*x
};
\addlegendentry{CM-like bound in \eqref{eq.CM-like_bound_asymptotic}}

\addplot[
    green!60!black,
    thick,
    dotted
] expression[
    domain=0:0.5,
    samples=200
]{
    (\r/(\r + 2)) - 2*(\r/(\r + 2))*x
};
\addlegendentry{Singleton-like bound in \eqref{eq.Singleton-like_bound_asymptotic}}

\addplot[
    purple,
    thick,
    dashdotted
] expression[
    domain=0:0.5,
    samples=200
]{
    (\r/(\r + 2)) - 2*(\r/(\r + 2))*((\q^(2*\r) - 1)/(\q^(2*\r) - \q^(2*\r - 2)))*x
};
\addlegendentry{Plotkin-like bound in \eqref{eq.Plotkin-like_bound_asymptotic}}

\end{axis}
\end{tikzpicture}
\end{figure}
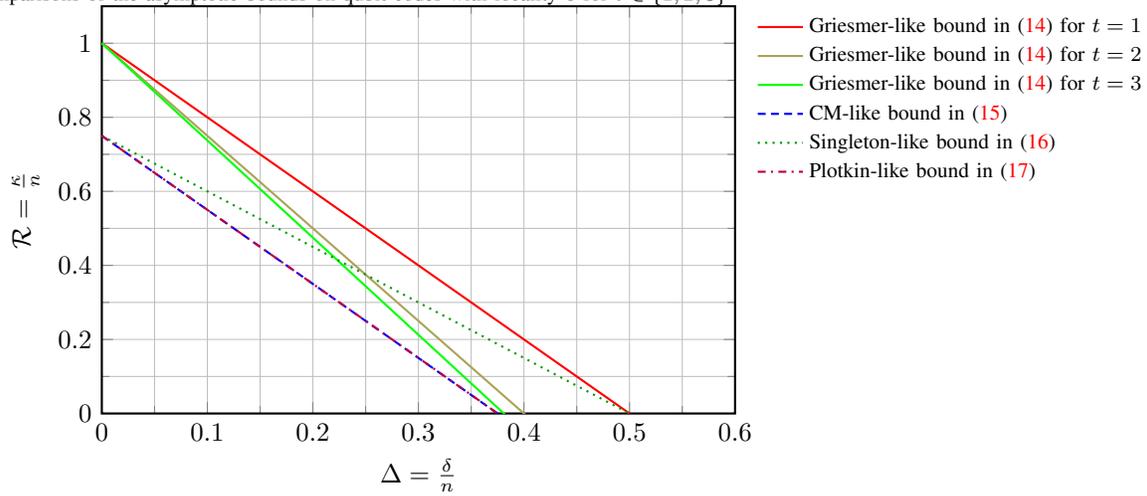

\begin{figure}[htbp]
  \caption{Comparisons of the asymptotic bounds on {qutrit codes} with locality 2 for $t\in \{1,2,3\}$}\label{fig.3}
\centering
\begin{tikzpicture}
\begin{axis}[
    width=10cm, height=7cm,
    xlabel={$ \Delta= \frac{\delta}{n}$},
    ylabel={$ \mathcal{R}=\frac{\kappa}{n} $},
    xmin=0, xmax=0.6,
    ymin=0, ymax=1.1,
    legend pos=north east,  
    grid=both,
    major grid style={line width=.2pt,draw=gray!50},
    minor tick num=1,
    thick,
    enlargelimits=false,
    legend style={
    font=\footnotesize,
    at={(1.02,1)},
    anchor=north west,
    align=left,
    draw=none,
    cells={anchor=west}}
]

\pgfmathsetmacro{\q}{3}
\pgfmathsetmacro{\r}{2}
\pgfmathsetmacro{\t}{1}
\pgfmathsetmacro{\tt}{2}
\pgfmathsetmacro{\ttt}{3}
\addplot[
    red,
    thick
] expression[
    domain=0:0.5,
    samples=200
]{
    1 - 2*((\q^(2*\t) - 1)/(\q^(2*\t) - \q^(2*\t - 2)))*x
};
\addlegendentry{Griesmer-like bound in \eqref{eq.Griesmer-like_bound_asymptotic} for $t=1$}

\addplot[
    yellow!60!black,
    thick
] expression[
    domain=0:0.5,
    samples=200
]{
    1 - 2*((\q^(2*\tt) - 1)/(\q^(2*\tt) - \q^(2*\tt - 2)))*x
};
\addlegendentry{Griesmer-like bound in \eqref{eq.Griesmer-like_bound_asymptotic} for $t=2$}

\addplot[
    green,
    thick
] expression[
    domain=0:0.5,
    samples=200
]{
    1 - 2*((\q^(2*\ttt) - 1)/(\q^(2*\ttt) - \q^(2*\ttt - 2)))*x
};
\addlegendentry{Griesmer-like bound in \eqref{eq.Griesmer-like_bound_asymptotic} for $t=3$}

\addplot[
    blue,
    thick,
  densely dashed,
] expression[
    domain=0:0.5,
    samples=200
]{
    (\r/(\r + 2)) - 2*(\r/(\r + 2))*(\q^2 / (\q^2 - 1))*x
};
\addlegendentry{CM-like bound in \eqref{eq.CM-like_bound_asymptotic}}

\addplot[
    green!60!black,
    thick,
    dotted
] expression[
    domain=0:0.5,
    samples=200
]{
    (\r/(\r + 2)) - 2*(\r/(\r + 2))*x
};
\addlegendentry{Singleton-like bound in \eqref{eq.Singleton-like_bound_asymptotic}}

\addplot[
    purple,
    thick,
    dashdotted
] expression[
    domain=0:0.5,
    samples=200
]{
    (\r/(\r + 2)) - 2*(\r/(\r + 2))*((\q^(2*\r) - 1)/(\q^(2*\r) - \q^(2*\r - 2)))*x
};
\addlegendentry{Plotkin-like bound in \eqref{eq.Plotkin-like_bound_asymptotic}}

\end{axis}
\end{tikzpicture}
\end{figure}
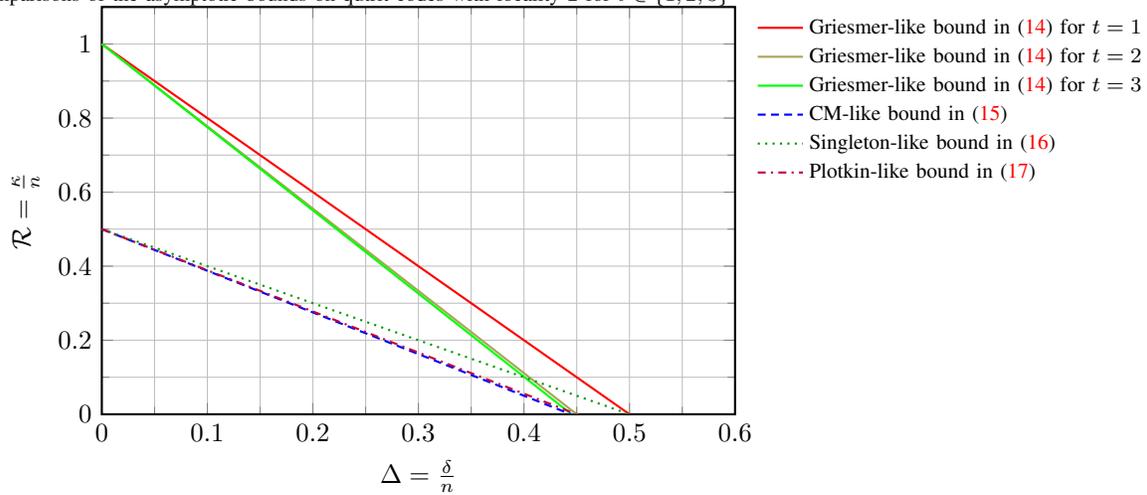

\section{Explicit constructions of qLRCs from the Hermitian construction}\label{sec.4}

According to Lemma \ref{lem.Hermitian_qLRC}, it is sufficient to {obtain} a Hermitian dual-containing cLRC 
for constructing a qLRC from the Hermitian construction. 
It has also been emphasized before that NMDS codes supporting $t$-designs are good candidates for obtaining optimal cLRCs. 
Therefore, if we can construct more new infinite families of NMDS codes supporting $t$-designs and determine 
{when they are Hermitian dual-containing}, then explicit constructions of good qLRCs can be deduced 
by using the Hermitian construction. 
This also offers meaningful contributions to design theory, given that most existing linear codes supporting designs are limited to fixed and small dimensions \cite{DT2022,DT2020,HW2023,HWL2023,XCQ2022,XCLW2024,XLX2025,DSY2024}.
The sequel is devoted to this purpose. 

\subsection{More NMDS codes with general or flexible dimensions supporting $t$-designs and optimal cLRCs}\label{sec.design}

In this subsection, we consider the family of Han-Zhang codes \cite{HZ2024} defined as follows 
and prove that their minimum weight codewords can support $t$-designs by using the subset sum theory. 
{Furthermore, we show that cLRCs derived from NMDS codes supporting $t$-designs 
are optimal with respect to four different bounds} presented in Lemma \ref{lem.bound_LRC} simultaneously, 
and produce four classes of such optimal cLRCs from Han-Zhang codes. 
In the following, we first recall the definition of Han-Zhang codes.

\begin{definition}
  Let $n$ and $k$ be positive integers such that $2\leq k\leq  n-1\leq q-1$. 
  Let $\SSS=\{a_1,a_2,\ldots,a_n\}\subseteq \F_q$ and $\vvv=(v_1,v_2,\ldots,v_n)\in (\F_q^*)^n$. 
  A {\em Han-Zhang code} $\HZ_k(\SSS,\vvv)$ is an $[n,k]_q$ linear code with a generator matrix 
\begin{align}\label{eq.M_k}
	G_{\HZ_k(\SSS,\vvv)} =\left[\begin{array}{cccc}
		v_1 & v_2 & \ldots & v_n  \\ 
    v_1a_1 & v_2a_2 & \ldots & v_na_n  \\ 
    \vdots & \vdots & \ddots & \vdots \\
    v_1a_1^{k-2} & v_2a_2^{k-2} & \ldots & v_na_n^{k-2}  \\
    v_1a_1^{k} & v_2a_2^{k} & \ldots & v_na_n^{k}
	\end{array}\right].   
\end{align}
\end{definition}

Let $\mathcal{F}\subseteq \F_q$ and $b\in \F_q$. 
The  {\em{subset sum problem}} over $\mathcal{F}$ is to determine 
if there is a subset $\emptyset \neq \{x_1,x_2,\ldots,x_{\ell}\}\subseteq \mathcal{F}$ such that 
\begin{align}\label{eq.subset sum problem def}
	x_1+x_2+\ldots+x_{\ell}=b.     
\end{align}
Generally, the subset sum problem is known to be NP-complete. 
Let $N({\ell},b,\mathcal{F})$ be the number of subsets $\{x_1,x_2,\ldots,x_{\ell}\}\subseteq \mathcal{F}$ such that \eqref{eq.subset sum problem def} holds. 
{Using a geometric approach}, Han and Zhang proved that any $[n,k]_q$ Han–Zhang code is either an MDS code or an NMDS code for $2 \leq k \leq n - 1 \leq q - 2$, 
depending on the corresponding subset sum problems \cite{HZ2024}. 
Furthermore, it is not hard to verify that such a result also holds for the case where $n=q$. 
In summary, we have the following result, which is a slight improvement of \cite[Proposition 2.5]{HZ2024}, as the case $n=q$ is included. 

\begin{lemma}\label{lem.parameters of Han-Zhang codes}
  Let $n$ and $k$ be positive integers such that $2\leq k\leq n-1\leq q-1$. 
  Let $\SSS=\{a_1,a_2,\ldots,a_n\}\subseteq \F_q$ and $\vvv=(v_1,v_2,\ldots,v_n)\in (\F_q^*)^n$. 
  Then the following statements hold. 
  \begin{itemize}
    \item [\rm 1)] $\HZ_k(\SSS,\vvv)$ is an $[n,k,n-k+1]_q$ MDS code if and only if $N(k,0,\SSS)=0$.
    \item [\rm 2)] $\HZ_k(\SSS,\vvv)$ is an $[n,k,n-k]_q$ NMDS code if and only if $N(k,0,\SSS)>0$.  
  \end{itemize}
  
\end{lemma}

In the sequel, we deduce $t$-designs from $\HZ_k(\SSS,\vvv)$. 
Specifically, we consider the following two cases:
\begin{itemize}
  \item $\SSS=\mathcal{A}\subseteq \F_{2^m}$ is an additive subgroup with $|\mathcal{A}|=2^{m_1}$ and $3\leq m_1\leq m$;
  \item $\SSS=\F_{2^{m_2}}^*\subseteq \F_{2^m}$ with $m_2\mid m$.
\end{itemize} 
To do this, we need some explicit formulas for $N(\ell,b,\SSS)$. 

\begin{lemma}{\em (\!\! \cite[Corollary 1.4]{LW2012})}\label{lem.ssp_additive_subgroup}
   Let $\mathcal{A}$ be any additive subgroup of $\F_q$ with $|\mathcal{A}|=n$. 
   For any $b\in \mathcal{A}$, the following statements hold.
   \begin{itemize}
    \item [\rm 1)] If $p\nmid k$, then 
    \begin{align}
      N(k,b,\mathcal{A})=\frac{1}{n}\binom{n}{k}.
    \end{align}

    \item [\rm 2)] If $p\mid k$, then 
    \begin{align}
      N(k,b,\mathcal{A})=\frac{1}{n}\binom{n}{k}+(-1)^{k+\frac{k}{p}}\frac{v(b)}{n}\binom{\frac{n}{p}}{\frac{k}{p}}~{\rm with}~
      v(b)=\left\{ 
        \begin{array}{ll}
          n-1, & \mbox{if}~b=0, \\ 
          -1, & \mbox{if}~b\neq 0. 
        \end{array}
      \right.
    \end{align}
   \end{itemize}
\end{lemma}

\begin{lemma}{\em (\!\! \cite[Theorem 1.2]{LW2008})}\label{lem.ssp_Fq*}
   We have that 
    \begin{align}
      N(k,b,\F_q^*)=\frac{1}{q}\binom{q-1}{k}+(-1)^{k+ \lfloor\frac{k}{p}\rfloor}\frac{v(b)}{q}\binom{\frac{q}{p}}{\lfloor\frac{k}{p}\rfloor}~{\rm with}~
      v(b)=\left\{ 
        \begin{array}{ll}
          q-1, & \mbox{if}~b=0, \\ 
          -1, & \mbox{if}~b\neq 0. 
        \end{array}
      \right.
    \end{align}
\end{lemma}

\subsubsection{\bf \em Infinite families of NMDS codes with flexible dimensions supporting $3$-designs}
~\\

Let $e(i,j,z)$ be the {\em counting function} for even integers, which returns the number of even integers in the set $\{i,j,z\}$, 
where $i,j,z$ are any integers. 
For examples, $e(1,0,3)=1$ and $e(2,5,6)=2$. 
Using this notation, we have the following result.

\begin{theorem}\label{th.333design}
  Let $\SSS=\mathcal{A}\subseteq \F_{2^m}$ be an additive subgroup with $|\mathcal{A}|=2^{m_1}$ 
  and $\vvv=(v_1,v_2,\ldots,v_{2^{m_1}})\in (\F_{2^m}^*)^{2^{m_1}}$, where $3\leq m_1\leq m$.   
	For each $3\leq k\leq 2^{m_1}-3$, the Han-Zhang code $\HZ_k(\mathcal{A},\vvv)$ is a $[2^{m_1},k,2^{m_1}-k]_{2^m}$ NMDS code.  
	Moreover, if $k$ is even, then the following statements hold. 
  \begin{itemize}
    \item [\rm 1)] The set of  minimum weight codewords of  $\HZ_k(\mathcal{A},\vvv)^{\perp}$ supports 
    a $3$-$(2^{m_1},k,\lambda_1^{\perp})$ design with 
    \begin{align}\label{eq.333design}
      \lambda_1^{\perp}=\sum_{i=0}^{k-3}\sum_{j=0}^{k-3-i}\sum_{z=0}^{k-3-i-j}{\Delta(k,i,j,z)},~{\rm and}~
      {\Delta(k,i,j,z)}=\left\{ 
        \begin{array}{ll}
          -\delta_1+(-1)^{\frac{k+2}{2}}(n-1)\delta_2, & \mbox{if}~e(i,j,z)=0, \\ 
          -\delta_1+(-1)^{\frac{k}{2}}\delta_2, & \mbox{if}~e(i,j,z)=2, \\
          \delta_1, & \mbox{else},  
        \end{array}
      \right.
    \end{align}
    where $$\delta_1=\frac{1}{2^{m_1}}\binom{2^{m_1}}{k-3-i-j-z},  ~{\rm and}~  \delta_2=\frac{1}{2^{m_1}}\binom{2^{m_1-1}}{\frac{k}{2}}.$$ 
    
    \item [\rm 2)] The set of minimum weight codewords of $\HZ_k(\mathcal{A},\vvv)$ supports 
    a $3$-$(2^{m_1},2^{m_1}-k,\lambda_1)$ design with 
    \begin{align}\label{eq.333design_lambda}
        \lambda_1=\frac{\lambda_1^{\perp}\binom{2^{m_1}-3}{k}}{\binom{2^{m_1}-3}{k-3}}
    \end{align}
  \end{itemize}
\end{theorem}
\begin{IEEEproof}
	Since $q=2^m\geq 8$ and $3\leq k\leq 2^{m_1}-3$, 
  one can easily check from Lemma \ref{lem.ssp_additive_subgroup} that $N(k,0,\mathcal{A})>0$ 
  (see also \cite[Corollary 1]{HF2023} or \cite[Corollary 2.8]{LW2008} for a similar proof). 
	It then follows from Lemma~\ref{lem.parameters of Han-Zhang codes}.2) 
  that $\HZ_k(\mathcal{A},\vvv)$ is a $[2^{m_1},k,2^{m_1}-k]_{2^m}$ NMDS code. 

  1) 
    Since $\HZ_k(\mathcal{A},\vvv)$ is NMDS, so is $\HZ_k(\mathcal{A},\vvv)^{\perp}$, which implies that $d(\HZ_k(\mathcal{A},\vvv)^{\perp})=k$, 
  where $k$ is even. 
	Let $\ccc=(c_1,c_2,\ldots,c_{2^{m_1}})\in \HZ_k(\mathcal{A},\vvv)^{\perp}$ with $\wt(\ccc)=k$ and $\supp(\ccc)=\{s_1,s_2,\dots,s_k\}$. 
	Hence, $c_{s_t}=u_{s_t}\in \F_q^*$ for $1\leq t\leq k$ and $c_v=0$ for all $v\in \{1,2,\ldots,2^{m_1}\}\setminus \{s_1,s_2,\ldots,s_k\}$. 
	Set $x_t=a_{s_t}$ and $v'_t=v_{s_t}$ for $1\leq t\leq k$, where 
  $a_{s_t}$ and $v_{s_t}$ are the $s_t$-th element in $\mathcal{A}=\{a_1,a_2,\ldots,a_{2^{m_1}}\}$ and $\vvv=(v_1,v_2,\ldots,v_{2^{m_1}})$, respectively.  
  Let 
  \begin{align*}
    M_{k,k}=\left[\begin{array}{cccc}
      v_1' & v_2' & \ldots & v_k'   \\ 
      v_1'x_1 & v_2'x_2 & \ldots &  v_k'x_{k}  \\ 
      \vdots & \vdots & \ddots  & \vdots \\
      v_1'x_1^{k-2} & v_2'x_2^{k-2} & \ldots  & v_k'x_{k}^{k-2}  \\
      v_1'x_1^{k} & v_2'x_2^{k} & \ldots  & v_k'x_{k}^{k}
    \end{array}\right].
  \end{align*}

Let ${\bf 0}$ be a zero vector of some appropriate length.  
	Since $\ccc\in \HZ_k(\mathcal{A},\vvv)^{\perp}$, we have 
	\begin{align}\label{eq.M_{k,k}u=0}
		M_{k,k}\uuu^T=\mathbf{0},
	\end{align}
	where $\uuu=(u_{s_1}, u_{s_2}, \ldots, u_{s_k})$. 
	Note that $\rank(M_{k,k})=k-1$, as $\uuu\neq {\bf 0}$ and 
	the first $k-1$ rows and $k-1$ columns of $M_{k,k}$ form a $(k-1)\times (k-1)$ Vandermonde matrix. 
  It then implies that the number of nonzero solutions  $\{u_{s_1}, u_{s_2}, \ldots, u_{s_k}\}\subseteq (\F_q^*)^k$ 
	of the system of equations (\ref{eq.M_{k,k}u=0}) equals $q-1$. 
	Furthermore, we deduce that all codewords of weight $k$ in $\HZ_k(\mathcal{A},\vvv)^{\perp}$ form 
	the set $\{a\ccc:~ a\in \F_q^*\}$ and all their supports are the set $\{s_1,s_2,\ldots,s_k\}$. 
	As a conclusion, each codeword of weight $k$ as well as its nonzero multiples in $\HZ_k(\mathcal{A},\vvv)^{\perp}$ 
	with support $\{s_1,s_2,\ldots,s_k\}$ corresponds to  the set $\{x_1,x_2,\ldots,x_k\}$.

  Then, to prove that the set of minimum weight codewords of $\HZ_k(\mathcal{A},\vvv)^{\perp}$ supports a $3$-design, 
  it suffices to prove that the number of choices of $\{x_4,x_5,\ldots,x_k\}$ such that $M_{k,k}$ has rank $k-1$ 
  is independent of any three fixed elements $x_1,x_2$, and $x_3$. 
  Note that 
  \begin{align*}
    \rank(M_{k,k})=k-1 & \Longleftrightarrow \det(M_{k,k})=\prod_{\ell=1}^{k}v_{\ell}'\prod_{1\leq \ell_1<\ell_2\leq k}{(x_{\ell_2}-x_{\ell_1})} (x_1+x_2+\ldots+x_k)=0\\ 
                       & \Longleftrightarrow x_1+x_2+\ldots+x_k=0 \\
                       & \Longleftrightarrow x_4+x_5+\ldots+x_k=x_1+x_2+x_3.  
  \end{align*}
  Therefore, the number of choices of $\{x_4,x_5,\ldots,x_k\}$ such that $M_{k,k}$ has rank $k-1$ 
  is equal to 
  \begin{align}\label{eq.ssp1}
    N(k-3,x_1+x_2+x_3,\mathcal{A}\setminus \{x_1,x_2,x_3\}),
  \end{align} 
  $i.e.,$ the number of solutions of the following subset sum problem
	\begin{align*}
		x_4+x_5+\ldots+x_k=x_1+x_2+x_3,~{\rm where}~\{x_4,x_5,\ldots,x_k\}\subseteq \mathcal{A}\setminus \{x_1,x_2,x_3\}.  
	\end{align*}

  By the inclusion-exclusion principle, by considering whether $x_3$ appears in the solution of \eqref{eq.ssp1}, 
  we have that 
  \begin{align*}
    & N(k-3,x_1+x_2+x_3,\mathcal{A}\setminus \{x_1,x_2,x_3\})\\
  = & N(k-3,x_1+x_2+x_3,\mathcal{A}\setminus \{x_1,x_2\}) - N(k-4,x_1+x_2+2x_3,\mathcal{A}\setminus \{x_1,x_2,x_3\}).
  \end{align*}
  Repeating the above process, we can further deduce that
	\begin{align}\label{eq.solution of ssp1}
		\begin{split}
			 & N(k-3,x_1+x_2+x_3,\mathcal{A}\setminus \{x_1,x_2,x_3\})  \\
			= & \sum_{i=0}^{k-3} (-1)^i N(k-3-i,x_1+x_2+(i+1)x_3,\mathcal{A}\setminus \{x_1,x_2\}) \\
			= & \sum_{i=0}^{k-3} (-1)^i \sum_{j=0}^{k-3-i} (-1)^j N(k-3-i-j,x_1+(j+1)x_2+(i+1)x_3,\mathcal{A}\setminus \{x_1\}) \\
      = & \sum_{i=0}^{k-3} (-1)^i \sum_{j=0}^{k-3-i} (-1)^j \sum_{z=0}^{k-3-i-j} (-1)^z N(k-3-i-j-z, (z+1)x_1+(j+1)x_2+(i+1)x_3,\mathcal{A}) \\
      = & \sum_{i=0}^{k-3} \sum_{j=0}^{k-3-i} \sum_{z=0}^{k-3-i-j} (-1)^{i+j+z} N(k-3-i-j-z, (z+1)x_1+(j+1)x_2+(i+1)x_3,\mathcal{A}).
		\end{split}
	\end{align} 
  Since $4\leq k\leq 2^{m_1}-4$ is even, we have the following cases described in Table \ref{tab.parity_ijz} 
  by considering the parity of $i,j,z$, where ``$o$'' and ``$e$'' denote odd and even integers, respectively. 
  \begin{table}[htbp]
    \centering
    \caption{Cases in terms of the parity of $i,j,z$}\label{tab.parity_ijz}       
    \vspace{-2mm}
    \scalebox{1}{
        \begin{tabular}{c|c|c|c|c}
     \hline 
     $\{i,j,z\}$  & $e(i,j,z)$ & $i+j+z$ & $k-3-i-j-z$ & $(z+1)x_1+(j+1)x_2+(i+1)x_3$ \\ \hline  \hline
      $(o,o,o)$ & $0$ & $o$ & $e$ & $0$ \\ \hline
      $(o,o,e)$ & $1$ & $e$ & $o$ & $x_1$ \\ \hline
      $(o,e,o)$ & $1$ & $e$ & $o$ & $x_2$ \\ \hline
      $(e,o,o)$ & $1$ & $e$ & $o$ & $x_3$ \\ \hline
      $(o,e,e)$ & $2$ & $o$ & $e$ & $x_1+x_2\neq 0$ \\ \hline
      $(e,o,e)$ & $2$ & $o$ & $e$ & $x_1+x_3\neq 0$ \\ \hline
      $(e,e,o)$ & $2$ & $o$ & $e$ & $x_2+x_3\neq 0$ \\ \hline
      $(e,e,e)$ & $3$ & $e$ & $o$ & $x_1+x_2+x_3$ \\ \hline
    \end{tabular}}
\end{table} 
Although we cannot directly determine whether $x_1,x_2,x_3$, and $x_1+x_2+x_3$ are $0$ in $\mathcal{A}$, 
the corresponding cases in Table \ref{tab.parity_ijz} imply that $2\nmid (k-3-i-j-z)$, 
and hence, we still know the exact values of $N(k-3-i-j-z, (z+1)x_1+(j+1)x_2+(i+1)x_3,\mathcal{A})$ for these cases. 

Furthermore, according to Table \ref{tab.parity_ijz} and Lemma \ref{lem.ssp_additive_subgroup}, we can immediately conclude that 
\begin{align}\label{eq.N(k-3-i-j-z)}
  \begin{split}
      & (-1)^{i+j+z} N(k-3-i-j-z, (z+1)x_1+(j+1)x_2+(i+1)x_3,\mathcal{A})\\ 
= & \left\{ 
        \begin{array}{ll}
          -\delta_1+(-1)^{\frac{k+2}{2}}(n-1)\delta_2, & \mbox{if}~e(i,j,z)=0, \\ 
          -\delta_1+(-1)^{\frac{k}{2}}\delta_2, & \mbox{if}~e(i,j,z)=2, \\
          \delta_1, & \mbox{else},  
        \end{array}
\right. 
  \end{split}
\end{align}
where $\delta_1=\frac{1}{2^{m_1}}\binom{2^{m_1}}{k-3-i-j-z}$ and 
$\delta_2=\frac{1}{2^{m_1}}\binom{\frac{2^{m_1}}{2}}{\frac{k}{2}}=\frac{1}{2^{m_1}}\binom{2^{m_1-1}}{\frac{k}{2}}.$
Noting that $\delta_1$ and $\delta_2$ are fixed for any given $i,j,z$, and $k$, 
it then turns out from \eqref{eq.N(k-3-i-j-z)} that 
the value of $(-1)^{i+j+z} N(k-3-i-j-z, (z+1)x_1+(j+1)x_2+(i+1)x_3,\mathcal{A})$ is only dependent on the values of $i,j,z$, and $k$, 
and hence, we can abbreviate it as ${\Delta(k,i,j,z)}$. 
In other words, for a given $k$,  
the number of choices of $\{x_4,x_5,\ldots,x_k\}$ such that $M_{k,k}$ has rank $k-1$ 
is independent of $x_1,x_2$, and $x_3$, and it is always equal to 
\begin{align*}
  \sum_{i=0}^{k-3}\sum_{j=0}^{k-3-i}\sum_{z=0}^{k-3-i-j}{\Delta(k,i,j,z)}.
\end{align*} 

Combining the above discussion, we now have that the set of codewords of weight $k$ in $\HZ_k(\mathcal{A},\vvv)^{\perp}$ supports 
a $3$-$(2^{m_1},k,\lambda_1^{\perp})$ design, where 
	$$\lambda_1^{\perp}=\sum_{i=0}^{k-3}\sum_{j=0}^{k-3-i}\sum_{z=0}^{k-3-i-j}{\Delta(k,i,j,z)}.$$
This competes the proof of 1).

2) It turns out from \eqref{eq.complementary_design_parameters} that the set of minimum weight codewords of $\HZ_k(\mathcal{A},\vvv)$ supports 
	a $3$-$(2^{m_1},2^{m_1}-k,\lambda_1)$ design with 
	$$\lambda_1=\frac{\lambda_1^{\perp}\binom{2^{m_1}-3}{k}}{\binom{2^{m_1}-3}{k-3}},$$
  where  $\lambda_1^{\perp}$ is the same as in 1). 
  This competes the proof of 2).
\end{IEEEproof}

\begin{theorem}\label{th.333design_wd}
  Let $\SSS=\mathcal{A}\subseteq \F_{2^m}$ be an additive subgroup with $|\mathcal{A}|=2^{m_1}$ 
  and $\vvv=(v_1,v_2,\ldots,v_{2^{m_1}})\in (\F_{2^m}^*)^{2^{m_1}}$, where $3\leq m_1\leq m$.   
	For any {even integer $k$} satisfying $4\leq k\leq 2^{m_1}-4$, the polynomial weight enumerators of 
  $\HZ_k(\mathcal{A},\vvv)$ and  $\HZ_k(\mathcal{A},\vvv)^{\perp}$ are given by 
	$$A(z)=1+\sum_{i=2^{m_1}-k}^{2^{m_1}}A_iz^i~{\rm and}~A^{\perp}(z)=1+\sum_{i=k}^{2^{m_1}}A^{\perp}_iz^i,$$  {respectively,} 
	where $A_{2^{m_1}-k}=A^{\perp}_k=\frac{\lambda_1^{\perp}2^{m_1+1}(2^{m}-1)(2^{m_1}-1)(2^{m_1-1}-1)}{k(k-1)(k-2)}$ 
  and $\lambda_1^{\perp}$ is given as in \eqref{eq.333design}. 
	Moreover, $A_i$ and $A^{\perp}_i$ are the same as those shown in Lemma \ref{lem.NMDS weight distribution}. 
\end{theorem}
\begin{IEEEproof}
	It follows from \eqref{eq.parameters of t-design}, Lemma \ref{lem.complement design}, and  Theorem \ref{th.333design} that 
	$$A_{2^{m_1}-k}=A^{\perp}_k=\frac{\lambda_1^{\perp}(2^m-1)\binom{2^{m_1}}{3}}{\binom{k}{3}}=\frac{\lambda_1^{\perp}2^{m_1+1}(2^{m}-1)(2^{m_1}-1)(2^{m_1-1}-1)}{k(k-1)(k-2)}.$$ 
	Then the desired results directly follow from Lemma \ref{lem.NMDS weight distribution}. 
\end{IEEEproof}

\subsubsection{\bf \em Infinite families of NMDS codes with general dimensions supporting $2$-designs}
~\\

For now, we consider $\SSS=\F_{2^{m_2}}^*\subseteq \F_{2^m}$ with $m_2\mid m$, 
and remove the even integer restriction on $k$ in Theorem \ref{th.333design}. 
We will get another infinite family of NMDS codes with general dimensions supporting $t$-designs. 

\begin{theorem}\label{th.222design}
  Let $\SSS=\F_{2^{m_2}}^*\subseteq \F_{2^m}$ with $m_2\mid m$ and $\vvv=(v_1,v_2,\ldots,v_{2^{m_2}-1})\in (\F_{2^m}^*)^{2^{m_2}-1}$.  
	For each $3\leq k\leq 2^{m_2}-4$, the Han-Zhang code $\HZ_k(\F_{2^{m_2}}^*,\vvv)$ is a $[2^{m_2}-1,k,2^{m_2}-1-k]_{2^m}$ NMDS code.  
	Moreover, the following statements hold. 
  \begin{itemize}
    \item [\rm 1)] The set of  minimum weight codewords of  $\HZ_k(\F_{2^{m_2}}^*,\vvv)^{\perp}$ supports 
    a $2$-$(2^{m_2}-1,k,\lambda_2^{\perp})$ design with 
    \begin{align}\label{eq.222design}
      \lambda_2^{\perp}=\sum_{i=0}^{k-2}\sum_{j=0}^{k-2-i} {\Delta'(k,i,j)},~{\rm and}~
      {\Delta'(k,i,j)}=\left\{ 
        \begin{array}{ll}
          \delta'_1+(-1)^{k+ \lfloor\frac{k-i-j}{2}\rfloor -1}(2^{m_2}-1)\delta'_2, & \mbox{if}~ij~{\rm is~odd}, \\ 
          (-1)^{i+j}\delta'_1+(-1)^{k+ \lfloor\frac{k-i-j}{2}\rfloor}\delta'_2, & \mbox{if}~ij~{\rm is~even}, 
        \end{array}
      \right.
    \end{align}
    where $$\delta'_1=\frac{1}{2^{m_2}}\binom{2^{m_2}-1}{k-2-i-j},  ~{\rm and}~  \delta'_2=\frac{1}{2^{m_2}}\binom{2^{m_2-1}-1}{\lfloor \frac{k-i-j}{2}\rfloor-1}.$$ 
    
    \item [\rm 2)] The set of minimum weight codewords of $\HZ_k(\F_{2^{m_2}}^*,\vvv)$ supports 
    a $2$-$(2^{m_2}-1,2^{m_2}-1-k,\lambda_2)$ design with 
    \begin{align}
          \lambda_2=\frac{\lambda_2^{\perp}\binom{2^{m_2}-3}{k}}{\binom{2^{m_2}-3}{k-2}}
    \end{align}
  \end{itemize}
\end{theorem}
\begin{IEEEproof}
	Since $q=2^m\geq 8$ and $3\leq k\leq 2^{m_2}-4$ with $m_2\mid m$, 
  one can easily check that $N(k,0,\F_{2^{m_2}}^*)>0$ (see also \cite[Corollary 2.7]{LW2008} for a similar proof). 
	It then follows from Lemma~\ref{lem.parameters of Han-Zhang codes}.2) 
  that $\HZ_k(\F_{2^{m_2}}^*,\vvv)$ is a $[2^{m_2},k,2^{m_2}-k]_{2^m}$ NMDS code. 

  1) 
  By a similar argument as in the proof of Theorem \ref{th.333design}.1), 
  to prove that the set of minimum weight codewords of $\HZ_k(\F_{2^{m_2}}^*,\vvv)^{\perp}$ supports a $2$-design, 
  it suffices to prove that the number of choices of $\{x_3,x_4,\ldots,x_k\}$ such that $M_{k,k}$ has rank $k-1$ 
  is independent of any two fixed elements $x_1$ and $x_2$. 
  Therefore, we consider $N(k-2,x_1+x_2,\F_{2^{m_2}}^*\setminus \{x_1,x_2\})$, $i.e.,$ the number of solutions of the following subset sum problem
	\begin{align*}
		x_3+x_4+\ldots+x_k=x_1+x_2,~{\rm where}~\{x_3,x_4,\ldots,x_k\}\subseteq \F_{2^{m_2}}^*\setminus \{x_1,x_2\}.  
	\end{align*}

  Following the same way as in \eqref{eq.solution of ssp1}, we have that 
	\begin{align}\label{eq.solution of ssp2}
		\begin{split}
			 & N(k-2,x_1+x_2,\F_{2^{m_2}}^*\setminus \{x_1,x_2\})  \\
			= & \sum_{i=0}^{k-2} (-1)^i N(k-2-i,x_1+(i+1)x_2,\F_{2^{m_2}}^*\setminus \{x_1\}) \\
			= & \sum_{i=0}^{k-2} (-1)^i \sum_{j=0}^{k-2-i} (-1)^j N(k-2-i-j,(j+1)x_1+(i+1)x_2,\F_{2^{m_2}}^*) \\
      = & \sum_{i=0}^{k-2} \sum_{j=0}^{k-2-i}  (-1)^{i+j} N(k-2-i-j, (j+1)x_1+(i+1)x_2,\F_{2^{m_2}}^*).
		\end{split}
	\end{align} 
  Since $x_1\neq x_2\in \F_{2^{m_2}}^*$, it is easy to check that $(j+1)x_1+(i+1)x_2=0$ if and only if $ij$ is odd. 
  It then follows from  Lemma \ref{lem.ssp_Fq*} that 
\begin{align}\label{eq.N(k-2-i-j)}
  \begin{split}
      & (-1)^{i+j} N(k-2-i-j, (j+1)x_1+(i+1)x_2,\F_{2^{m_2}}^*)\\ 
= & \left\{ 
        \begin{array}{ll}
          \delta'_1+(-1)^{k+ \lfloor\frac{k-i-j}{2}\rfloor -1}(2^{m_2}-1)\delta'_2, & \mbox{if}~ij~{\rm is~odd}, \\ 
          (-1)^{i+j}\delta'_1+(-1)^{k+ \lfloor\frac{k-i-j}{2}\rfloor}\delta'_2, & \mbox{if}~ij~{\rm is~even}, 
        \end{array}
      \right.
  \end{split}
\end{align}
where $\delta'_1=\frac{1}{2^{m_2}}\binom{2^{m_2}-1}{k-2-i-j}$ and $\delta'_2=\frac{1}{2^{m_2}}\binom{2^{m_2-1}-1}{\lfloor \frac{k-i-j}{2}\rfloor-1}.$ 
Again, we deduce that the value of $(-1)^{i+j} N(k-2-i-j, (j+1)x_1+(i+1)x_2,\F_{2^{m_2}}^*)$ is only dependent on the values of $i,j$, and $k$, 
and hence, we can abbreviate it as ${\Delta'(k,i,j)}$. 
This competes the proof of 1).

2) It turns out from \eqref{eq.complementary_design_parameters} that the set of minimum weight codewords of $\HZ_k(\F_{2^{m_2}}^*,\vvv)$ supports 
	a $2$-$(2^{m_2}-1,2^{m_2}-1-k,\lambda_2)$ design with 
	$$\lambda_2=\frac{\lambda_2^{\perp}\binom{2^{m_2}-3}{k}}{\binom{2^{m_2}-3}{k-2}},$$
  where  $\lambda_2^{\perp}$ is the same as in 1). 
  This competes the proof of 2).
\end{IEEEproof}

\begin{theorem}\label{th.222design_wd}
  Let $\SSS=\F_{2^{m_2}}^*\subseteq \F_{2^m}$ with $m_2\mid m$ and $\vvv=(v_1,v_2,\ldots,v_{2^{m_2}-1})\in (\F_{2^m}^*)^{2^{m_2}-1}$.  
	For each $3\leq k\leq 2^{m_2}-4$, the polynomial weight enumerators of 
  $\HZ_k(\F_{2^{m_2}}^*,\vvv)$ and  $\HZ_k(\F_{2^{m_2}}^*,\vvv)^{\perp}$ are given by 
	$$A(z)=1+\sum_{i=2^{m_2}-k-1}^{2^{m_2}-1}A_iz^i~{\rm and}~A^{\perp}(z)=1+\sum_{i=k}^{2^{m_2}-1}A^{\perp}_iz^i,$$  {respectively,} 
	where $A_{2^{m_2}-k-1}=A^{\perp}_k=\frac{\lambda_2^{\perp}(2^{m}-1)(2^{m_2}-1)(2^{m_2}-2)}{k(k-1)}$ 
  and $\lambda_2^{\perp}$ is given as in \eqref{eq.222design}.
	Moreover, $A_i$ and $A^{\perp}_i$ are the same as those shown in Lemma \ref{lem.NMDS weight distribution}. 
\end{theorem}
\begin{IEEEproof}
	It follows from \eqref{eq.parameters of t-design}, Lemma \ref{lem.complement design}, and  Theorem \ref{th.222design} that 
	$$A_{2^{m_2}-k-1}=A^{\perp}_k=\frac{\lambda_2^{\perp}(2^{m}-1)\binom{2^{m_2}-1}{2}}{\binom{k}{2}}=\frac{\lambda_2^{\perp}(2^{m}-1)(2^{m_2}-1)(2^{m_2}-2)}{k(k-1)}.$$ 
	Then the desired results directly follow from Lemma \ref{lem.NMDS weight distribution}.  
\end{IEEEproof}

\begin{remark}\label{rem.designs}
  Let ${\bf 1}$ be an all-one vector of an appropriate length. 
  {Heng and Wang conjectured in \cite[Conjecture 36]{HW2023} that 
  the Han-Zhang code $\HZ_k(\F_{2^m}^*,{\bf 1})$ is an NMDS code and its minimum weight codewords support a $2$-design for each $3\leq k\leq 2^m-4$.} 
  Using the subset sum theory, Li $et~al.$ \cite{LZM2024} proved this conjecture. 
  According to \cite{HW2023}, it is the first infinite family of NMDS codes with general dimensions supporting $t$-designs for $t\geq 2$. 
  Based on a different method, Zhang $et~al.$ \cite{ZZWL2025} proved this conjecture again, and further proved that the minimum weight codewords of $\HZ_k(\F_{2^m},{\bf 1})$ 
  support a $3$-design, {where $k$ is an even integer satisfying $4\leq k\leq 2^m-4$}.

  In this section, we improve the method used in \cite{LZM2024} to prove that the minimum weight codewords of $\HZ_k(\SSS,{\bf v})$ support $3$-designs and $2$-designs,  
  for $\SSS$ being any additive subgroup of $\F_{2^m}$ and $\SSS=\F_{2^{m_2}}^*$ with $m_2\mid m$, respectively, where $\vvv$ is an arbitrary vector in $(\F_{2^m}^*)^{|\SSS|}$. 
  We emphasize that our results not only generalize those in \cite{LZM2024} and \cite{ZZWL2025},
  but also enable the incorporation of Hermitian dual-containing structures into Han-Zhang codes (see Subsection~\ref{sub.qLRCs} for more details),
  which is crucial for constructing good qLRCs via the Hermitian construction. 
\end{remark}

\subsubsection{\bf \em Applications to optimal cLRCs}
~\\

In this subsection, we consider the applications of 
the above Han-Zhang codes supporting $t$-designs 
to optimal cLRCs. 
First of all, we describe a result on optimal cLRCs from NMDS codes supporting $t$-designs. 
To this end, we recall a relationship between cLRCs and linear codes supporting $t$-designs \cite{TFDTZ2023}. 


\begin{lemma}{\rm (\!\! \cite[Corollary 3]{TFDTZ2023})}\label{lem.locality}
	Let $\C$ be an $[n,k]_q$ linear code. If $d^{\perp}=d(\C^{\perp})\geq 2$ and  
  $(\mathcal{P}(\C^{\perp}),\BBB_{d^{\perp}}(\C^{\perp}))$ is 
	a $1$-$(n,d^{\perp},\lambda^{\perp})$ design with $\lambda^{\perp}\geq 1$, then $\C$ has locality $d^{\perp}-1$. 
\end{lemma}

As we {mentioned} in Subsection \ref{sec1.3}, NMDS codes supporting $t$-designs yield optimal cLRCs with respect to the Singleton-like bound and the CM bound \cite{TFDTZ2023}. 
In the following, we give a full explanation for this claim and show that such cLRCs are also optimal with respect to the other two bounds {presented} in Lemma \ref{lem.bound_LRC}.

\begin{theorem}\label{th.optimal_LRCs_from_designs}
    Let $\C$ be an $[n,k,n-k]_q$ NMDS code with $n-k\geq 2$ and the minimum weight codewords of its dual 
    supports a $1$-$(n,k,\lambda^{\perp})$ design with $\lambda^{\perp}\geq 1$. 
    Then $\C$ is an optimal $(n,k,n-k,q;k-1)$ cLRC with respect to the 
    Griesmer-like bound, the CM bound, the Singleton-like bound, and the Plotkin-like bound 
    given in Lemma \ref{lem.bound_LRC}.   
\end{theorem}
\begin{IEEEproof}
    Under the given assumption, it follows from Lemma \ref{lem.locality} that 
    $\C$ has locality $d(\C^{\perp})-1=k-1$, $i.e.,$ 
    $\C$ is an $(n,k,n-k,q;k-1)$ cLRC. 
    Since $n-k\geq 2$, the optimality of $\C$ with respect to the CM bound and the Singleton-like bound 
    can be derived from \cite[Theorem 15]{TFDTZ2023} directly. 
    Noting also that $\left\lceil \frac{k}{r} \right\rceil=\left\lceil \frac{k}{k-1} \right\rceil=2$, we have 
    \begin{align*}
         \max_{1\leq \ell\leq \left\lceil \frac{k}{r} \right\rceil-1}
                                                                \left\{\ell(r+1)+\sum_{i=0}^{k-\ell r-1}\left\lceil \frac{d}{q^i} \right\rceil \right\} 
      =  (k-1)+1+\sum_{i=0}^{k-(k-1)-1}\left\lceil \frac{n-k}{q^i} \right\rceil  
      =  n                                                       
    \end{align*}
    and 
     \begin{align*}
         \min_{1\leq \ell\leq \left\lceil \frac{k}{r} \right\rceil-1}
                                                                \left\{\frac{q^{k-\ell r-1}(q-1)(n-\ell(r+1))}{q^{k-\ell r}-1} \right\} 
      =  \frac{q^{k-(k-1)-1}(q-1)(n-((k-1)+1))}{q^{k-(k-1)}-1}  
      =  n-k.                                                        
    \end{align*}
    Combining with the length, minimum distance, and Lemmas \ref{lem.bound_LRC}.1) and 4), we immediately deduce that 
    $\C$ is also optimal with respect to the Griesmer-like bound and the Plotkin-like bound. 
    This completes the proof. 
\end{IEEEproof}

Now, we have four families of cLRCs derived from Han-Zhang codes that are simultaneously optimal with respect to all the four bounds.

\begin{theorem}\label{th.optimal_LRC111}
Let $\SSS=\mathcal{A}\subseteq \F_{2^m}$ be an additive subgroup with $|\mathcal{A}|=2^{m_1}$ 
  and $\vvv=(v_1,v_2,\ldots,v_{2^{m_1}})\in (\F_{2^m}^*)^{2^{m_1}}$, where $3\leq m_1\leq m$.   
	For any {even integer $k$} satisfying $4\leq k\leq 2^{m_1}-4$, the following results hold. 
  \begin{itemize}
    \item [\rm 1)] $\HZ_k(\mathcal{A},\vvv)$ is a $(2^{m_1},k,2^{m_1}-k,{2^m};k-1)$ optimal cLRC 
    with respect to the Griesmer-like bound, the CM bound, the Singleton-like bound, and the Plotkin-like bound 
    given in Lemma \ref{lem.bound_LRC}, simultaneously.    

    \item [\rm 2)] $\HZ_k(\mathcal{A},\vvv)^{\perp}$ is a $(2^{m_1},2^{m_1}-k,k,{2^m};2^{m_1}-k-1)$ optimal cLRC 
    with respect to the Griesmer-like bound, the CM bound, the Singleton-like bound, and the Plotkin-like bound 
    given in Lemma \ref{lem.bound_LRC}, simultaneously.    
  \end{itemize}  
\end{theorem}
\begin{IEEEproof}
  We only prove 1) while 2) can be proved in a similar way. 
  It follows from Theorem \ref{th.333design}.2) that $\HZ_k(\mathcal{A},\vvv)$ is a 
  $[2^{m_1},k,2^{m_1}-k]_{2^m}$ NMDS code whose minimum weight codewords support a $3$-$(2^{m_1},2^{m_1}-k,\lambda_1)$ design, 
  where $\lambda_1\geq 1$ is defined as in \eqref{eq.333design_lambda}. 
  {It implies that} the set of minimum weight codewords of $\HZ_k(\mathcal{A},\vvv)$ can also support 
  a $1$-$(2^{m_1},2^{m_1}-k,\lambda_1')$ design with $\lambda_1'\geq 1$ (see \cite[Theorem 4.4]{DT2022}). 
  Since $2^{m_1}-k\geq 2$ under the given assumption on $k$,  
  1) follows easily from Theorem \ref{th.optimal_LRCs_from_designs}. 
  This completes the proof.
\end{IEEEproof}

\begin{theorem}\label{th.optimal_LRC222}
  Let $\SSS=\F_{2^{m_2}}^*\subseteq \F_{2^m}$ with $m_2\mid m$ and $\vvv=(v_1,v_2,\ldots,v_{2^{m_2}-1})\in (\F_{2^m}^*)^{2^{m_2}-1}$.  
	For each $3\leq k\leq 2^{m_2}-4$, the following results hold. 
  \begin{itemize}
    \item [\rm 1)]   $\HZ_k(\F_{2^{m_2}}^*,\vvv)$ is a $(2^{m_2}-1,k,2^{m_2}-k-1,{2^m};k-1)$ optimal cLRC 
    with respect to the Griesmer-like bound, the CM bound, the Singleton-like bound, and the Plotkin-like bound 
    given in Lemma \ref{lem.bound_LRC}, simultaneously.  
    
    \item [\rm 2)]  $\HZ_k(\F_{2^{m_2}}^*,\vvv)^{\perp}$ is a $(2^{m_2}-1,2^{m_2}-k-1,k,{2^m};2^{m_2}-k-2)$ optimal cLRC 
    with respect to the Griesmer-like bound, the CM bound, the Singleton-like bound, and the Plotkin-like bound 
    given in Lemma \ref{lem.bound_LRC}, simultaneously.   
  \end{itemize}
\end{theorem}
\begin{IEEEproof}
  The proof is similar to that of Theorem \ref{th.optimal_LRC111}, and hence, we omit it here.
\end{IEEEproof}

\subsection{Three new families of optimal qLRCs}\label{sub.qLRCs}

Now, we present three new families of qLRCs derived from Han–Zhang codes 
supporting $t$-designs obtained in Subsection~\ref{sec.design}, 
using the Hermitian construction in Lemma~\ref{lem.Hermitian_qLRC}. 
We also determine their optimality by comparing their parameters with the bounds stated in Theorem~\ref{th.qLRCs_bounds}, 
and refer to Remark \ref{rem.optimal_qLRCs} for the definition of optimal qLRCs. 
As a result, we indeed provide positive and nontrivial answers to Open Problem \ref{prob.qLRCs_from_Hermitian}. 
To do this, we note the following fact. 

According to \cite{LX2004}, for any set $\SSS=\{a_1,a_2,\ldots,a_{n}\}\subseteq \F_{q^2}$,    
any vector $\vvv=(v_1,v_2,\ldots,v_{n})\in (\F_{q^2}^*)^{n}$, and a positive integer $k$ satisfying 
$1\leq k+1\leq n\leq q^2$, there is an $[n,k+1,n-k]_{q^2}$ 
{\em generalized Reed-Solomon (GRS) code}, denoted by $\GRS_k(\SSS,\vvv)$, with the generator matrix 
\begin{align*}
    G_{\GRS_{k+1}(\SSS,\vvv)}=\left[\begin{array}{cccc}
      v_1 & v_2 & \ldots & v_n \\
      v_1a_1 & v_2a_2 & \ldots & v_na_n \\
      \vdots & \vdots & \ddots & \vdots \\
      v_1a_1^{k-2} & v_2a_2^{k-2} & \ldots & v_na_n^{k-2} \\
      v_1a_1^{k-1} & v_2a_2^{k-1} & \ldots & v_na_n^{k-1} \\
      v_1a_1^{k} & v_2a_2^{k} & \ldots & v_na_n^{k}
    \end{array}\right].
\end{align*}
Combining with \eqref{eq.M_k}, we deduce that 
\begin{align}\label{eq.nested_codes}
  \HZ_{k}(\SSS,\vvv)\subseteq \GRS_{k+1}(\SSS,\vvv). 
\end{align}

Based on the key fact presented in \eqref{eq.nested_codes} and {the fact that $\C^{\perp}$ and $\C^{\perp_{\rm H}}$ have the same locality for any $q^2$-ary linear code $\C$}, 
we derive three new families of optimal qLRCs from Han-Zhang codes 
supporting $t$-designs and the Hermitian construction.

\begin{theorem}\label{th.qLRC111}
    Let $m$ and $m_1$ be two positive integers satisfying $m+3 \leq m_1\leq 2m$. 
    For any {even integer $k$} satisfying $4\leq  k\leq \left\lfloor \frac{2^{m_1}+2^m-1}{2^m+1} \right\rfloor-1$, 
    there exists a $[[2^{m_1},2^{m_1}-2k,\geq k]]_{2^m}$ qLRC $\mathcal{Q}_1$ with locality $2^{m_1}-k-1$,   
    which is optimal with respect to the Singleton-like bound in \eqref{eq.Singleton-like_bound} or the pure Singleton-like bound in \eqref{eq.Singleton-like_bound_pure}.  
\end{theorem}
\begin{IEEEproof}
From $m+3\leq m_1\leq 2m$, 
we verify that $\left\lfloor \frac{2^{m_1}+2^m-1}{2^m+1} \right\rfloor\geq 5$ and $m_1\geq 3$. 
Since $2^{m_1}\mid 2^{2m}$, we can take $\SSS=\mathcal{A}$ as a certain additive subgroup of $\F_{2^{2m}}$ 
with $|\mathcal{A}|=2^{m_1}$. 
According to \cite[Theorem 4.5]{QCWL2024}, there exists a vector 
$\vvv=(v_1,v_2,\ldots,v_{2^{m_1}})\in (\F_{2^{2m}}^*)^{2^{m_1}}$ such that 
$\GRS_{k+1}(\SSS,\vvv)$ is a $[2^{m_1},k+1,2^{m_1}-k]_{2^{2m}}$ Hermitian self-orthogonal MDS code 
for any $5\leq k+1\leq \left\lfloor \frac{2^{m_1}+2^m-1}{2^m+1} \right\rfloor$. 
With \eqref{eq.nested_codes}, we then deduce that 
\begin{align*}
    (\HZ_{k}(\mathcal{A},\vvv)^{\perp_{\rm H}})^{\perp_{\rm H}}=\HZ_{k}(\mathcal{A},\vvv)\subseteq \GRS_{k+1}(\mathcal{A},\vvv)\subseteq  
    \GRS_{k+1}(\mathcal{A},\vvv)^{\perp_{\rm H}}\subseteq \HZ_{k}(\mathcal{A},\vvv)^{\perp_{\rm H}}, 
\end{align*}
and hence, $\HZ_{k}(\mathcal{A},\vvv)^{\perp_{\rm H}}$ is Hermitian dual-containing. 

Since $\mathcal{A}$ is an additive subgroup of $\F_{2^{2m}}$, it further follows from 
Theorem \ref{th.optimal_LRC111} that $\HZ_{k}(\mathcal{A},\vvv)^{\perp_{\rm H}}$ is a 
$[2^{m_1},2^{m_1}-k,k]_{2^{2m}}$ Hermitian dual-containing NMDS code with locality $2^{m_1}-k-1$ 
for any {even integer $k$} satisfying $4\leq  k\leq \left\lfloor \frac{2^{m_1}+2^m-1}{2^m+1} \right\rfloor-1$, 
{as $\HZ_{k}(\mathcal{A},\vvv)^{\perp}$ and $\HZ_{k}(\mathcal{A},\vvv)^{\perp_{\rm H}}$ have the same locality}. 
Applying Lemma \ref{lem.Hermitian_qLRC} to $\HZ_{k}(\mathcal{A},\vvv)^{\perp_{\rm H}}$, 
we immediately obtain a $[[2^{m_1},2^{m_1}-2k,\delta\geq k]]_{2^m}$ qLRC $\mathcal{Q}_1$ with locality $2^{m_1}-k-1$. 

Next, we consider the optimality of $\mathcal{Q}_1$. 
Substituting the parameters of $\mathcal{Q}_1$ into the right of 
the Singleton-like bound given in \eqref{eq.Singleton-like_bound}, we have that 
\begin{align*}
    n-\kappa-2\left\lceil \frac{\kappa}{r} \right\rceil+4
    = 2^{m_1}-(2^{m_1}-2k)-2\left\lceil \frac{2^{m_1}-2k}{2^{m_1}-k-1} \right\rceil+4=2k+2, 
\end{align*}
which implies that the optimal minimum distance of $[[2^{m_1},2^{m_1}-2k,\delta]]_{2^m}$ qLRCs 
with locality $2^{m_1}-k-1$ is less than or equal to $k+1$, that is $\delta=k$ or $k+1$. 
In addition, substituting the parameters of $\mathcal{Q}_1$ into the right of 
the pure Singleton-like bound given in \eqref{eq.Singleton-like_bound_pure} yields that 
\begin{align*}
  n-\kappa-2\left\lceil \frac{n+\kappa}{2r} \right\rceil+4
    = 2^{m_1}-(2^{m_1}-2k)-2\left\lceil \frac{2^{m_1}+(2^{m_1}-2k)}{2\cdot (2^{m_1}-k-1)} \right\rceil+4=2k. 
\end{align*}
Therefore, $\mathcal{Q}_1$ is optimal with respect to the bound in \eqref{eq.Singleton-like_bound} if $\delta=k+1$, 
and with respect to the bound in \eqref{eq.Singleton-like_bound_pure} if $\delta=k$. 
This completes the proof. 
\end{IEEEproof}

\begin{theorem}\label{th.qLRC222}
    Let $m$ and $m_1$ be two positive integers satisfying $4\leq m_1\leq m$. 
    For any {even integer $k$} satisfying $4\leq  k\leq 2^{m_1-1}-1$, 
    there exists a $[[2^{m_1},2^{m_1}-2k,\geq k]]_{2^m}$ qLRC $\mathcal{Q}_2$ with locality $2^{m_1}-k-1$, 
    which is optimal with respect to the Singleton-like bound in \eqref{eq.Singleton-like_bound} or the pure Singleton-like bound in \eqref{eq.Singleton-like_bound_pure}.  
\end{theorem}
\begin{IEEEproof}
    Since $4\leq m_1\leq m$, we have $2^{m_1-1}\geq 8>5$ and can take $\SSS=\mathcal{A}$ 
    as a certain additive subgroup of $\F_{2^{m}}\subseteq \F_{2^{2m}}$. 
    With \cite[Theorem 3.5]{FFLZ2020}, there always exists a vector 
    $\vvv=(v_1,v_2,\ldots,v_{2^{m_1}})\in (\F_{2^{2m}}^*)^{2^{m_1}}$ such that 
    $\GRS_{k+1}(\SSS,\vvv)$ is a $[2^{m_1},k+1,2^{m_1}-k]_{2^{2m}}$ Hermitian self-orthogonal MDS code 
    for any $5\leq k+1\leq \left\lfloor \frac{2^{m_1}}{2} \right\rfloor=2^{m_1-1}$. 
    Using \eqref{eq.nested_codes} and Lemma \ref{lem.Hermitian_qLRC}, 
    we then deduce a $[[2^{m_1},2^{m_1}-2k,\delta\geq k]]_{2^m}$ qLRC $\mathcal{Q}_2$ with locality $2^{m_1}-k-1$ 
    for any {even integer $k$} satisfying $4\leq k\leq 2^{m_1}-1$ by an argument similar to that of Theorem \ref{th.qLRC111}. 
    Comparing the parameters of $\mathcal{Q}_2$ with the Singleton-like bound and the pure Singleton-like bound for qLRCs,  
    respectively given in \eqref{eq.Singleton-like_bound} and \eqref{eq.Singleton-like_bound_pure}, 
    we arrive at the claim of the optimality of $\mathcal{Q}_2$. 
    This completes the proof. 
\end{IEEEproof}

\begin{theorem}\label{th.qLRC333}
    Let $m$ and $m_2$ be two positive integers satisfying $4\leq m_2\leq m$ and $m_2\mid m$. 
    For any $4\leq  k\leq 2^{m_2-1}-1$, 
    there exists a $[[2^{m_2}-1,2^{m_2}-2k-1,\geq k]]_{2^m}$ qLRC $\mathcal{Q}_3$ with locality $2^{m_2}-k-2$, 
    which is optimal with respect to the Singleton-like bound in \eqref{eq.Singleton-like_bound} or the pure Singleton-like bound in \eqref{eq.Singleton-like_bound_pure}.  
\end{theorem}
\begin{IEEEproof}
    The proof is very similar to that of Theorem \ref{th.qLRC222} 
    and the main difference is that we use Theorem \ref{th.optimal_LRC222} instead of Theorem \ref{th.optimal_LRC111}. 
    We omit the details here.
\end{IEEEproof}

\begin{remark}{\em (Comparisons of qLRCs)}\label{rem.comparisons_qLRCs}
Currently, there are still relatively few explicit families of qLRCs constructed in the literature. 
Moreover, only four of these families are known to be optimal. 
In Table~\ref{tab:qLRCs}, we summarize all known families of $[[n,\kappa,\delta]]_{q}$ qLRCs with locality $r$, along with the three constructions introduced in this paper.
From the table, it is evident that our optimal constructions are new, 
as they exhibit {distinct localities}, variable dimensions, and larger minimum distances.
In addition, the notation ``$*$'' denotes that the corresponding family of qLRCs is optimal.
\begin{table*}[ht!]
\centering
\caption{{{Known families of $[[n,\kappa,\delta]]_q$ qLRCs with locality $r$}}}
\label{tab:qLRCs}
\setlength{\tabcolsep}{3pt}
\renewcommand{\arraystretch}{1.2}
\resizebox{\textwidth}{!}{
\begin{threeparttable}
\begin{tabular}{l|c|c|c|l}
\toprule
No. & $[[n,\kappa,\delta]]_q$  & $r$ & Condition & Reference\\ 
\midrule
1 & $[[7^{2^{m-1}}\cdot 2^{2^{m-1}-1},k,\delta]]_2$ & $2^{m+1}-1$ & $k\geq 2^{2^{m-1}-1}$, $\delta\geq 3$, and $m\geq 2$ & \cite[Example 30]{BGL2025} \\

2$^*$ & $[[n,2k-n,n-k+1]]_q$ & $k$ & there is an $[n,k]_{q^2}$ Hermitian dual-containing MDS code & \cite[Proposition 34]{GHMM2024} \\

3 & $[[q-1,(2\ell-q)\left(1-2/r\right)+\epsilon,\delta]]_{q}$ & $r\mid (q-1)$~{\rm and}~$r\geq 3$  & $-2\leq \epsilon\leq 2$,~$\ell\leq q$, and $\delta\geq \Delta_1\tnote{1}$ for prime $r$  & \cite[Definition 3.1]{GG-QLRC2025} \\ 

4$^*$ & $[[u(r+1),\kappa,\delta]]_{q}$ & $r\leq q-1$~{\rm and}~$r>2\delta+u-4$ & $\kappa$ and $\delta$ are determined by certain specific cases & \cite[Table \Rmnum{1}]{LCEL2025} \\

5 & $[[n,2k-n,\delta]]_q$ & $(r+1)\mid n$ & $n\leq q$, $\frac{n}{2}<k\leq \frac{nr}{r+1}$, and $\delta\geq \Delta_2\tnote{2}$ & \cite[Theorem 1]{SRT2025} \\

6$^*$ & $[[q^2,q^2-6,3]]_{q}$ & $q(q-1)-1$ & $q\geq 7$ & \cite[Theorem 6]{XZS2025} \\

7$^*$ & $[[q(q-1)/{h},q(q-1)/{h}-6,3]]_{q}$ & ${q(q-h-1)}/{h}-1$ & $q\geq 7$ & \cite[Theorem 9]{XZS2025} \\
\midrule
8$^*$ & $[[2^{m_1},2^{m_1}-2k,\geq k]]_{2^m}$ & $2^{m_1}-k-1$ & even $k$ with $4\leq k\leq \left\lfloor \frac{2^{m_1}+2^m-1}{2^m+1} \right\rfloor-1$ and $m+3\leq m_1\leq 2m$ & Theorem \ref{th.qLRC111} \\

9$^*$ & $[[2^{m_1},2^{m_1}-2k,\geq k]]_{2^m}$ & $2^{m_1}-k-1$ & even $k$ with $4\leq k\leq 2^{m_1-1}-1$ and $4\leq m_1\leq m$ & Theorem \ref{th.qLRC222} \\

10$^*$ & $[[2^{m_2}-1,2^{m_2}-2k-1,\geq k]]_{2^m}$ & $2^{m_2}-k-2$ & $4\leq k\leq 2^{m_2-1}-1$ and $4\leq m_2\leq m$, $m_2\mid m$ & Theorem \ref{th.qLRC333} \\
\bottomrule
\end{tabular}
\begin{tablenotes}
\footnotesize
\item[1] $\Delta_1=(q-1)\left(1-\frac{1}{2r}-\sqrt{\frac{1}{4r^2}+\frac{r-1}{r}\cdot \frac{\ell-1}{q-1}}\right)$ for prime $r$ was given in \cite[Theorem 3.1]{GG-QLRC2025}. 
A “folded” version of this family of qLRCs with smaller lengths was also given in \cite[Definition 3.2]{GG-QLRC2025}.
\item[2] $\Delta_2=\max\{r+1,n-\ell\}$, where the exact value of $\ell$ was determined by specific cases in \cite[Equation (3)]{SRT2025}.  
In addition, the {construction} is also dependent on the existence of the so-called “good” polynomials.
\end{tablenotes}
\end{threeparttable}
}
\end{table*}
\end{remark}

\section{Conclusion}\label{sec.5}

{Motivated by Open Problem \ref{prob.qLRCs_from_Hermitian} proposed in \cite{LCEL2025}, this paper studied bounds and explicit constructions of qLRCs 
using the Hermitian construction presented in Lemma \ref{lem.Hermitian_qLRC}.  
We established the Griesmer-like bound, the CM-like bound, 
the Singleton-like bound, and the Plotkin-like bound for qLRCs in Theorem \ref{th.qLRCs_bounds}, 
where our Singleton-like bound improved upon the one proposed in \cite{GHMM2024} for impure qLRCs.  
We also derived the corresponding asymptotic formulas in~\eqref{eq.Griesmer-like_bound_asymptotic}–\eqref{eq.Plotkin-like_bound_asymptotic}, 
which were then used to compare the tightness of these bounds. 
Motivated by Theorem \ref{th.optimal_LRCs_from_designs}, which showed that NMDS codes supporting $t$-designs yield optimal cLRCs, 
we employed NMDS Han-Zhang codes as the classical foundation for constructing qLRCs. 
In Theorems \ref{th.333design} and \ref{th.222design}, we proved that the minimum weight codewords of 
NMDS Han-Zhang codes and their duals, with flexible or general dimensions, can support $t$-designs for $t=3$ and $2$, respectively. 
By determining when such Han-Zhang codes are Hermitian dual-containing, 
we derived three new families of optimal qLRCs in Theorems \ref{th.qLRC111}, \ref{th.qLRC222}, and \ref{th.qLRC333}. 
We also confirmed that our qLRCs offered more flexible parameters than those obtained from the CSS construction 
in Remark \ref{rem.comparisons_qLRCs}.  
As a result, we provided positive and nontrivial answers to Open Problem \ref{prob.qLRCs_from_Hermitian}.
In addition, we obtained the polynomial weight enumerators of these Han-Zhang codes 
and four families of related optimal cLRCs in Theorems \ref{th.333design_wd}, \ref{th.222design_wd}, \ref{th.optimal_LRC111}, and \ref{th.optimal_LRC222}.


There are also several research directions that warrant further exploration. 
In the proof of Theorem \ref{th.optimal_LRC111}, we explained that our qLRCs are optimal with respect to 
the Singleton-like bound in \eqref{eq.Singleton-like_bound} if they are impure,  
or the pure Singleton-like bound in \eqref{eq.Singleton-like_bound_pure} otherwise. 
Note also that the authors of \cite{LCEL2025} proposed another open problem: 
{\em are there impure qLRCs that can meet the Singleton-like bound in \eqref{eq.Singleton-like_bound}?} 
Therefore, it is of particular interest to investigate the purity of our qLRCs. 
If they are impure, we not only derived three new families of optimal qLRCs, 
but also provided an affirmative answer to the aforementioned open problem.
If they are pure, we still obtained three new families of optimal qLRCs with flexible parameters.
However, in that case, alternative approaches or different code families may be needed to further address the problem in \cite{LCEL2025}.}

\end{sloppypar}

\begin{thebibliography}{99}


  \bibitem{ABHMT2018} A. Agarwal, A. Barg, S. Hu, A. Mazumdar, and I. Tamo, ``Combinatorial alphabet-dependent bounds for locally recoverable codes," IEEE Trans. Inf. Theory, vol. 64, no. 5, pp. 3481-3492, May 2018.
  \bibitem{BGL2025} K. Bu, W. Gu, and X. Li, ``Quantum locally recoverable code with intersecting recovery sets,"  ArXiv preprint. [Online] Available: \url{https://arxiv.org/pdf/2501.10354}, 2025.
	\bibitem{CM2013} V. Cadambe and A. Mazumdar, ``An upper bound on the size of locally recoverable codes,"  in Proc. 2013 Int. Sym. Network Coding (NetCod), Calgary, AB, Canada, pp. 1-5, 2013.
  \bibitem{CWLX2021} H. Chen, J. Weng, W. Luo, and L. Xu, ``Long optimal and small-defect LRC codes with unbounded minimum distances," IEEE Trans. Inf. Theory, vol. 67, no. 5, pp. 2786-2792, May 2021. 
  \bibitem{CQLLXZ2024} H. Chen, L. Qu, C. Li, S. Lyu, L. Xu, and M. Zhou, ``Generalized Singleton type upper bounds," IEEE Trans. Inf. Theory, vol. 70, no. 5, pp. 3298-3308, May 2024.


  \bibitem{DSY2024} C. Ding, Z. Sun, and Q. Yan, ``The support designs of several families of lifted linear codes," Des. Codes Cryptogr., vol. 93, pp. 1575–1595, Jun. 2025. 
  \bibitem{DT2020} C. Ding and C. Tang, ``Infinite families of near MDS codes holding $t$-designs,” IEEE Trans. Inf. Theory, vol. 66, no. 9,  pp. 5419-5428, Sep. 2020. 
  \bibitem{DT2022} C. Ding and C. Tang, {\em Designs From Linear Codes (Second Edition)}. World Scientific, 2021.

	\bibitem{DL1995-NMDS} S. Dodunekov and I. Landgev, ``On near-MDS codes,” J. Geom., vol. 54, no. 1–2, pp. 30–43, Nov. 1995.



	\bibitem{FW1997-sAMDS} A. Faldum and W. Willems, ``Codes of small defect," Des. Codes Cryptogr., vol. 10, no. 3, pp. 341-350, Mar. 1997. 
  \bibitem{FFLZ2020} W. Fang, F.-W. Fu, L. Li, and S. Zhu, ``Euclidean and Hermitian hulls of MDS codes and their applications to EAQECCs," IEEE Trans. Inf. Theory, vol. 66, no. 6, pp. 3527-3537, Jun. 2020.
  \bibitem{FTFCX2024} W. Fang, R. Tao, F.-W. Fu, B. Chen, and S.-T. Xia, ``Bounds and constructions of Singleton-optimal locally repairable codes with small localities," IEEE Trans. Inf. Theory, vol. 70, no. 10, pp. 6842-6856, Oct. 2024.


  \bibitem{GHMM2024} C. Galindo, F. Hernando, H. Mart\'in-Cruz, and R. Matsumoto, ``Quantum $(r,\delta)$-locally recoverable codes,” ArXiv preprint. [Online] Available: \url{https://arxiv.org/pdf/2412.16590}, 2024. 
  \bibitem{GG-QLRC2025} L. Golowic and V. Guruswami, ``Quantum locally recoverable codes," in Proc. Annu. ACM-SIAM Symp. Discrete Algorithms (SODA), Society for Industrial and Applied Mathematics, pp. 5512-5522, 2025. 
	\bibitem{GHSY2012} P. Gopalan, C. Huang, H. Simitci, and S. Yekhanin, ``On the locality of codeword symbols,” IEEE Trans. Inf. Theory, vol. 58, no. 11, pp. 6925–6934, Nov. 2012. 
  \bibitem{GFWH2019} M. Grezet, R. Freij-Hollanti, T. Westerb\"ack, and C. Hollanti, ``Alphabet-dependent bounds for linear locally repairable codes based on residual codes," IEEE Trans. Inf. Theory, vol. 65, no. 10, pp. 6089-6100, Oct. 2019.
  \bibitem{GJR2023} A. Gruica, B. Jany, and A. Ravagnani, "Duality and LP bounds for codes with locality," in Proc. 2023 IEEE Inf. Theory Workshop (ITW), Saint-Malo, France, pp. 347-352, 2023.

  \bibitem{HF2023} D. Han and C. Fan, ``Roth–Lempel NMDS codes of non-elliptic-curve type,” IEEE Trans. Inf. Theory, vol. 69, no. 9, pp. 5670-5675, Sep. 2023.
  \bibitem{HZ2024} D. Han and H. Zhang, ``Explicit constructions of NMDS self-dual codes,” Des. Codes Cryptogr., vol. 92, no. 11, pp. 3573–3585, Jul. 2024.
  \bibitem{HXSCFY2020} J. Hao, S.-T. Xia, K.W. Shum, B. Chen, F.-W. Fu, and Y. Yang, ``Bounds and constructions of locally repairable codes: parity-check matrix approach," IEEE Trans. Inf. Theory, vol. 66, no. 12, pp. 7465-7474, Dec. 2020.
  \bibitem{HW2023} Z. Heng and X. Wang, ``New infinite families of near MDS codes holding $t$-designs,” Discrete Math., vol. 346, no. 10, Art. no. 113538, Oct. 2023.
  \bibitem{HWL2023} Z. Heng, X. Wang, and X. Li, ``Constructions of cyclic codes and extended primitive cyclic codes with their applications,”  Finite Fields Appl., vol. 89, Art. no. 102208, Aug. 2023.
  \bibitem{HYM2025} Z. Heng, M. Yang, and Y. Ming, ``A family of self-orthogonal divisible codes with locality $2$," Discrete Math., vol. 348, no. 10, Art. no. 114529, Oct. 2025.
  \bibitem{HEHBANS2016} K. Heshami, D.G. England, P.C. Humphreys, P.J. Bustard, V.M. Acosta, J. Nunn, and B.J. Sussman, ``Quantum memories: Emerging applications and recent advances,” J. Modern Opt., vol. 63, no. 20, pp. 2005–2028, 2016.


  \bibitem{KKKS2006} A. Ketkar, A. Klappenecker, S. Kumar, and P. K. Sarvepalli, ``Nonbinary stabilizer codes over finite fields,” IEEE Trans. Inf. Theory, vol. 52, no. 11, pp. 4892–4914, Nov. 2006.
  \bibitem{KNF2019} S. Kruglik, K. Nazirkhanova, and A. Frolov, ``New bounds and generalizations of locally recoverable codes with availability," IEEE Trans. Inf. Theory, vol. 65, no. 7, pp. 4156-4166, Jul. 2019.
  \bibitem{LW2008} J. Li and D. Wan,  ``On the subset sum problem over finite fields,” Finite Fields Appl., vol. 14, no. 4, pp. 911–929, Nov. 2008.
  \bibitem{LW2012} J. Li and D. Wan,  ``Counting subset sums of finite abelian groups,” J. Comb. Theory A, vol. 119, no. 1, pp. 170–182, Aug. 2011.
  \bibitem{LZM2024} Y. Li, S. Zhu, and E. Mart\'inez-Moro, ``On $\ell$-MDS codes and a conjecture on infinite families of 1-MDS codes," IEEE Trans. Inf. Theory, vol. 70, no. 10, pp. 6899-6911, Oct. 2024.
  \bibitem{LX2004} S. Ling and C.~Xing, \emph{Coding Theory: A First Course}. Cambridge Univ. Press, Cambridge, 2004.
  \bibitem{LMSDL2023} C. Liu, M. Wang, S.A. Stein, Y. Ding, and A. Li, ``Quantum memory: A missing piece in quantum computing units,” ArXiv preprint. [Online] Available: \url{https://arxiv.org/pdf/2309.14432}, 2023.
  \bibitem{LCEL2025} G. Luo, B. Chen, M.F. Ezerman, and S. Ling, ``Bounds and constructions of quantum locally recoverable codes from quantum CSS codes,” IEEE Trans. Inf. Theory, vol. 71, no. 3, pp. 1794-1802, Mar. 2025.
  \bibitem{LEL2023} G. Luo, M. F. Ezerman, and S. Ling, ``Three new constructions of optimal locally repairable codes from matrix-product codes," IEEE Trans. Inf. Theory, vol. 69, no. 1, pp. 75-85, Jan. 2023.

  \bibitem{MRELW2023} L. Meßner, E. Robertson, L. Esguerra, K. L\"udge, and J. Wolters, ``Multiplexed random-access optical memory in warm cesium vapor,” Opt. Express, vol. 31, no. 6, pp. 10150–10158, Mar. 2023.

  \bibitem{NKV2024} A.K. Nayak, E. Chitambar, and L.R. Varshney, ``Reliable quantum memories with unreliable components,” Phys. Rev. A, vol. 110, no. 3, Art. no. 32423, Sep. 2024.
  \bibitem{NC2011} M.A. Nielsen and I.L. Chuang, ``{\em Quantum Computation and Quantum Information: 10th Anniversary Edition}", Cambridge University Press, 2011.

  \bibitem{P2025} B. Pacifico, ``Introducing locality in some generalized AG codes," Des. Codes Cryptogr., 2025. [Online] Available: \url{https://doi.org/10.1007/s10623-025-01597-w}.
  \bibitem{PKLK2012} N. Prakash, G.M. Kamath, V. Lalitha, and P.V. Kumar, ``Optimal linear codes with a local-error-correction property,” in Proc. IEEE Int. Symp. Inf. Theory (ISIT), Cambridge, MA, USA, pp. 2776–2780, 2012.


  \bibitem{QCWL2024} L. Qian, X. Cao, X. Wu, and W. Lu,``MDS codes with $l$-Galois hulls of arbitrary dimensions,” Des. Codes  Cryptogr., vol. 92, no. 7, pp. 1879–1902, Mar. 2024.

  \bibitem{S2025} K. Senthoor, ``Entanglement cost of erasure correction in quantum MDS codes," ArXiv preprint. [Online] Available: \url{https://arxiv.org/pdf/2505.20284}, 2025.
  \bibitem{SRT2025} S. Sharma, V. Ramkumar and I. Tamo, ``Quantum locally recoverable codes via good polynomials," IEEE J. Sel. Areas Inf. Theory, vol. 6, pp. 100-110, May 2025.


  \bibitem{TB2014} I. Tamo and A. Barg, ``A family of optimal locally recoverable codes," IEEE Trans. Inf. Theory, vol. 60, no. 8, pp. 4661-4676, Aug. 2014. 
  \bibitem{TFDTZ2023} P. Tan, C. Fan, C. Ding, C. Tang, and Z. Zhou, ``The minimum locality of linear codes,” Des. Codes  Cryptogr., vol. 91, no. 1, pp. 83–114, Aug. 2022.

  \bibitem{XZS2025} D. Xie, S. Zhu, and Z. Sun, ``Two families of optimal quantum locally recoverable codes,” Int. J. Theor. Phys., vol. 64, no. 4, pp. 1-17, Mar. 2025.
  \bibitem{XCQ2022} G. Xu, X. Cao, and L. Qu, ``Infinite families of 3-designs and 2-designs from almost MDS codes,” IEEE Trans. Inf.  Theory, vol. 68, no. 7, pp. 4344-4353, Jul. 2022.  
  \bibitem{XCLW2024} G. Xu, X. Cao, G. Luo, and H. Wu, ``Infinite families of 3-designs from special symmetric polynomials,” Designs, Codes  Cryptogr., vol. 92, no. 12, pp. 4487-4509, Dec. 2024.
  \bibitem{XLX2025} G. Xu, J. Li, and H. Xu, ``An infinite family of AMDS codes holding 2-designs,” Adv. Math. Commun., 2025. [Online] Available: \url{https://doi.org/10.3934/amc.2025031}.

  \bibitem{ZZWL2025} Y. Zhang, D. Zheng, X. Wang, and W. Lu, ``Several new infinite families of NMDS codes with arbitrary dimensions supporting $t$-designs,"  ArXiv preprint. [Online] Available: \url{https://arxiv.org/pdf/2504.06546}, 2025.
\end{thebibliography}
\end{document}